\documentclass[12pt]{imsart}
\usepackage[utf8]{inputenc}
\usepackage[letterpaper,  margin=1.2in]{geometry}
\usepackage{amsthm,amsmath}
\usepackage{ctable}
\usepackage{multirow}
\def\bfA{{\boldsymbol{A}}}
\def\bfB{{\boldsymbol{B}}}
\def\bfg{{\boldsymbol{g}}}  \def\bfG{{\boldsymbol{G}}}
\def\bfh{{\boldsymbol{h}}}
\def\bfH{{\boldsymbol{H}}}
\def\bfH{{\boldsymbol{H}}}
  \def\bfI{{\boldsymbol{I}}}
  \def\bfJ{{\boldsymbol{J}}}
  
  \def\bfQ{{\boldsymbol{Q}}}
\def\bfs{{\boldsymbol{s}}}  \def\bfS{{\boldsymbol{S}}}
\def\bfu{{\boldsymbol{u}}}  \def\bfU{{\boldsymbol{U}}}
\def\bfv{{\boldsymbol{v}}} \def\bfV{{\boldsymbol{V}}}
  \def\bfW{{\boldsymbol{W}}}
  \def\bfX{{\boldsymbol{X}}}
 \def\bfZ{{\boldsymbol{Z}}}

\def\bfOmega{{\boldsymbol{\Omega}}}
\def\bfSigma{{\boldsymbol{\Sigma}}}
   
\def\boldbeta{{\mbox{\boldmath $\beta$}}}     
  
\def\boldlambda{{\mbox{\boldmath $\lambda$}}} 
\def\boldtheta{{\mbox{\boldmath $\theta$}}}   

\def\ACov{\text{ACov}}

\def\ACov{\mbox{ACov}}

\def\be{\begin{eqnarray}}
\def\ee{\end{eqnarray}}

\date{\today}
\begin{document}
\title{Improving Estimation Efficiency for Two-Phase, Outcome-Dependent Sampling Studies}
\runtitle{Improving Estimation Efficiency for Two-Phase Studies}
%\author{Menglu Che}
\author{\fnms{Menglu} \snm{Che},}\footnote{Department of Biostatistics, School of Public Health, Yale University}
%\address{Department of Biostatistics, School of Public Health, Yale University}
\author{\fnms{Peisong} \snm{Han}}\footnote{Department of Biostatistics, School of Public Health, University of Michigan}
%\address{Department of Biostatistics, School of Public Health, University of Michigan}
%\and
\author{and \fnms{Jerald F.} \snm{Lawless}}
\footnote{Department of Statistics and Actuarial Science, University of Waterloo}
\runauthor{Che, Han and Lawless}
\vspace{1em}
\begin{abstract}
    Two-phase outcome dependent sampling (ODS) is widely used in many
fields, especially when certain covariates are expensive and/or difficult
to measure. For two-phase ODS, the conditional maximum likelihood
(CML) method is very attractive because it can handle
zero Phase 2 selection probabilities and avoids modeling the covariate distribution.
However, most existing CML-based methods use only the Phase 2 sample
and thus may be less efficient than other methods. We propose
a general empirical likelihood method that uses CML augmented with additional  information in the whole Phase 1 sample to improve estimation
efficiency. The proposed method maintains the ability to handle zero
selection probabilities and avoids modeling the covariate distribution,
but can lead to substantial efficiency gains over CML  in the inexpensive covariates, or in the influential covariate when a surrogate is available, because of an effective use of the Phase 1 data. Simulations and a real data illustration using NHANES data are presented.
\end{abstract}
\maketitle
\section{Introduction}

For many large scale studies, some covariates may be expensive or
difficult to measure, in which case cost-efficient two-phase sampling
designs are desired for data collection and analysis. In Phase 1,
measurements on the outcome and the less expensive covariates are taken from a random sample of individuals, and then in Phase 2 a sub-sample
of Phase 1 individuals is selected and their measurements on the expensive
or difficult-to-measure covariates are taken. In order to improve
the estimation efficiency in model fitting, the Phase 2 sampling often
depends on the outcome values observed in Phase 1, known as outcome-dependent
sampling (ODS). Some well-known examples of ODS include case-control
and case-cohort studies (e.g., \cite{breslow1997maximum,scott1997fitting,kulich2004improving,kim2013more,schaid2013two,keogh2014case}) for rare outcomes and two-phase studies stratified by a continuous outcome (e.g., \cite{zhou2002semiparametric,weaver2005estimated,barnett2013detecting,lin2013quantitative,bjornland2018powerful,espin2018two}). In many applications, for example, \cite{kulich2004improving,schaid2013two} and \cite{espin2018two}, an inexpensive covariate is a surrogate or inaccurate measurement for an expensive covariate, and is not part of the outcome model of interest. We refer to this as a surrogate covariate scenario.

To fix notation, let $Y$ denote the outcome, $\boldsymbol{X}$ the
less expensive or easy-to-measure covariates, and $\boldsymbol{Z}$
the expensive or difficult-to-measure covariates. The primary objective
is to estimate the parameter $\boldsymbol{\beta}$ in a model $f(Y|\boldsymbol{X},\boldsymbol{Z};\boldsymbol{\beta})$
for the conditional distribution $f(Y|\boldsymbol{X},\boldsymbol{Z})$, or in the surrogate covariate case, $f(Y|\boldsymbol{Z};\boldsymbol{\beta})$
for the conditional distribution $f(Y|\boldsymbol{Z})$. We view the latter as a special case and use the former in our theoretical derivations hereafter. 
Two-phase designs collect data on $Y$ and $\boldsymbol{X}$
in Phase 1 and $\boldsymbol{Z}$ in Phase 2 and can substantially
reduce the study cost. Let $R$ denote the indicator for an individual
from Phase 1 to be selected for Phase 2. For ODS the selection depends
on the value of $Y$; let $\pi(Y,\boldsymbol{X})=P(R=1|Y,\boldsymbol{X})$
denote the selection probability, where we allow
$\pi(Y,\boldsymbol{X})$ to possibly depend on $\boldsymbol{X}$ as
well. The observed data from two-phase designs are then a random sample
$(Y_{i},\boldsymbol{X}_{i},R_{i},R_{i}\boldsymbol{Z}_{i})$, $i=1,\ldots,n$,
of $n$ individuals. Thus, the two-phase ODS essentially is a missing
data problem where the covariates $\boldsymbol{Z}$ are missing at
random (MAR) \cite{little2019statistical}. 

There have been many methods developed for analyzing data from two-phase ODS studies. The full maximum likelihood (ML) method
models the conditional distribution $f(\boldsymbol{Z}|\boldsymbol{X})$,
either parametrically or nonparametrically, in addition to the model
of interest $f(Y|\boldsymbol{X},\boldsymbol{Z};\boldsymbol{\beta})$,
and then maximizes the joint likelihood (e.g., \cite{lawless1999semiparametric,zhang2006semiparametric})
\[
\prod_{i=1}^{n}\{f(Y_{i}|\boldsymbol{X}_{i},\boldsymbol{Z}_{i};\boldsymbol{\beta})f(\boldsymbol{Z}_{i}|\boldsymbol{X}_{i})\}^{R_{i}}\{\int f(Y_{i}|\boldsymbol{X}_{i},\boldsymbol{Z};\boldsymbol{\beta})f(\boldsymbol{Z}|\boldsymbol{X}_{i})d\boldsymbol{Z}\}^{1-R_{i}}.
\]
The inverse probability weighting (IPW) method \cite{horvitz1952generalization} solves 
\[
\sum_{i=1}^{n}\frac{R_{i}}{\pi(Y_{i},\boldsymbol{X}_{i})}\boldsymbol{s}_{\boldsymbol{\beta}}(Y_{i},\boldsymbol{X}_{i},\boldsymbol{Z}_{i};\boldsymbol{\beta})={\bf 0},
\]
where $\boldsymbol{s}_{\boldsymbol{\beta}}(Y,\boldsymbol{X},\boldsymbol{Z};\boldsymbol{\beta})$
is the score function $\partial\log\{f(Y|\boldsymbol{X},\boldsymbol{Z};\boldsymbol{\beta})\}/\partial\boldsymbol{\beta}^T$.
The IPW method can be extended following the augmented IPW (AIPW)
method \cite{robins1994estimation,qin2009empirical} to improve the estimation efficiency.
The conditional maximum likelihood (CML) method maximizes the likelihood
\begin{equation}
\prod_{i=1}^{n}f_{c}(Y_{i}|\boldsymbol{X}_{i},\boldsymbol{Z}_{i},R_{i}=1;\boldsymbol{\beta})^{R_{i}},\label{CML lik}
\end{equation}
where 
\begin{align}
f_{c}(Y|\boldsymbol{X},\boldsymbol{Z},R=1;\boldsymbol{\beta})=\frac{f(Y|\boldsymbol{X},\boldsymbol{Z};\boldsymbol{\beta})\pi(Y,\boldsymbol{X})}{\int f(Y|\boldsymbol{X},\boldsymbol{Z};\boldsymbol{\beta})\pi(Y,\boldsymbol{X})dY}\label{condden}
\end{align}
is the conditional distribution of $Y$ for individuals chosen for Phase 2 (e.g., \cite{breslow1988logistic,breslow1997weighted,scott1997fitting,scott2011fitting}).
Some authors also proposed methods based on solving score equations
with certain components replaced by empirical estimates
(e.g., \cite{lawless1999semiparametric,chatterjee2003pseudoscore}). 

The CML method is particularly attractive in many situations. Compared
to the FML method, it does not need to model $f(\boldsymbol{Z}|\boldsymbol{X})$,
which is not of primary interest. For FML, when $f(\boldsymbol{Z}|\boldsymbol{X})$
is incorrectly modeled, the resulting estimator is no longer consistent.
Modeling $f(\boldsymbol{Z}|\boldsymbol{X})$ is particularly challenging
for two reasons. First, it is a missing data problem and the model
fitting cannot be done based on a complete case analysis due to the
dependence of the selection on the outcome $Y$. Second, even postulating
a model for $f(\boldsymbol{Z}|\boldsymbol{X})$ can be difficult when
both $\boldsymbol{Z}$ and $\boldsymbol{X}$ are multi-dimensional 
and/or of mixed types. Compared to the IPW/AIPW method which can also avoid modelling the covariate distribution, the CML method
can deal with a wider class of two-phase ODS designs. In particular,
it can be applied to studies where some individuals by design have
a zero probability of entering Phase 2, in which case the IPW/AIPW
method is no longer applicable. Two-phase ODS with possibly zero selection
probabilities is widely used in practice. For instance, in genome-wide
association studies (GWAS) where sequencing everyone may be  expensive,
researchers often only sample persons with extreme values of a quantitative trait
 (e.g., \cite{barnett2013detecting,lin2013quantitative,bjornland2018powerful}). In certain
situations, the optimal two-phase design may involve zero selection
probabilities for some individuals \cite{tao2017efficient}. 

However, the CML method discards the information
from Phase 1 individuals that do not enter Phase 2. Some authors have
proposed alternative versions of CML to make use of the Phase 1 data
to improve estimation efficiency (e.g., \cite{lawless1999semiparametric,scott2011fitting,che2020empirical,rivera2020augmented}), but most of these
proposals are either for special settings, such as binary outcomes,
or make use of Phase 1 data only through a model for the selection
probability. An approach that applies in a broad range of settings and uses Phase 1 data to improve on the CML method is highly desirable. 

We propose a general and effective procedure to improve on the CML
method. Our method can deal with zero
selection probabilities provided that the selection probabilities are mutually independent. The proposed method summarizes the Phase 1 data
information through a model for $f(Y|\boldsymbol{X})$. When all Phase
1 individuals have positive probabilities of being selected to enter
Phase 2, the model for $f(Y|\boldsymbol{X})$ is a working model and
the proposed method has a higher efficiency than the CML method, regardless of the correctness of this model. When some Phase 1 individuals have
a zero probability of entering Phase 2, a correct specification of
the model for $f(Y|\boldsymbol{X})$ is needed in theory to have an
efficiency improvement. However, in either case, $f(Y|\boldsymbol{X})$
is typically easy to model because $Y$ is a scalar and the specification
and fitting of this model is done based on fully observed Phase 1
data, and thus model diagnostic techniques can be applied. Therefore,
even though the model for $f(Y|\boldsymbol{X})$ may be subject to
misspecification, especially because it needs to be mathematically compatible with $f(Y|\boldsymbol{X},\boldsymbol{Z};\boldsymbol{\beta})$ to be correct, 
the model misspecification should be mild if appropriate diagnostics
are used. The proposed method can deal with different types of outcomes,
including both binary outcomes and continuous outcomes. %The efficiency gain is mostly in the inexpensive covariates, namely $\boldbeta_X$, which is suitable for the cases where the inexpensive covariate is the most predictive, such as the data provided in our real data illustration. 

The rest of the paper is organized as follows. In Section 2, we consider the case where the selection probability of every individual into Phase 2 is positive. In Section 3 we make the extension to the case where some individuals have zero selection probability into Phase 2. Section 4 provides simulation results concerning both binary and continuous outcomes, positive and zero selection probabilities, expensive and surrogate covariates. Section 5 contains an illustration of our method using real data from the 
 National Health and Nutrition Examination Survey (NHANES) 2015-2016. Some discussions are given in Section 6. Some mathematical details and other materials are provided in the Appendix. 

\section{The Method Under Positive Selection Probabilities}

To make good use of the Phase 1 data to improve the estimation efficiency,
we specify a working model $f(Y|\boldsymbol{X};\boldsymbol{\theta})$
for $f(Y|\boldsymbol{X})$ indexed by parameter $\boldsymbol{\theta}$.
Let $\boldsymbol{h}(Y,\boldsymbol{X};\boldsymbol{\theta})=\partial\log\{f(Y|\boldsymbol{X};\boldsymbol{\theta})\}/\partial\boldsymbol{\theta}^T$
denote the corresponding score function and $\widehat{\boldsymbol{\theta}}$
the solution to the score equation $\sum_{i=1}^{n}\{\boldsymbol{h}(Y_{i},\boldsymbol{X}_{i};\boldsymbol{\theta})\}={\bf 0}$
based on the fully observed Phase 1 data. In addition, let $\boldsymbol{\theta}^{*}$
denote the probability limit of $\widehat{\boldsymbol{\theta}}$ when
$n\to\infty$. Then the information contained in the Phase 1 data
can be summarized in the form
\begin{equation}
E\{\boldsymbol{h}(Y,\boldsymbol{X};\boldsymbol{\theta}^{\ast})\}={\bf 0}.\label{phase1_infor}
\end{equation}
The $\boldsymbol{\theta}^{*}$ minimizes the Kullback-Leibler distance
between the family of distributions $f(Y|\boldsymbol{X};\boldsymbol{\theta})$
indexed by $\boldsymbol{\theta}$ and the true distribution $f(Y|\boldsymbol{X})$.
In many two-phase ODS studies the Phase 1 sample size is large and
only a small subset of subjects are selected to enter Phase 2. In
this case the uncertainty associated with $\widehat{\boldsymbol{\theta}}$
estimated using Phase 1 data may be ignored compared to the uncertainty
associated with Phase 2 data and thus $\boldsymbol{\theta}^{*}$ may
be treated as equal to $\widehat{\boldsymbol{\theta}}$ for CML-based
methods. Our method can deal with both this case and the case where the Phase 1 sample size is not large such that its uncertainty is not ignorable. For the method in this section $f(Y|\boldsymbol{X};\boldsymbol{\theta})$
is a working model and does not need to be correctly specified.

\subsection{Using the known selection probability }%$\pi(Y,\boldsymbol{X})$}

With all Phase 1 individuals having positive probabilities, $\pi(Y,\boldsymbol{X})>0$,
of entering Phase 2, in this subsection we consider estimators that
directly use the known values of $\pi(Y,\boldsymbol{X})$. These values
are known by design for two-phase ODS. We first transform the information
in (\ref{phase1_infor}) so it can be incorporated into the CML-based
estimation. Note that 
\begin{align*}
{\bf 0} & =E\{\boldsymbol{h}(Y,\boldsymbol{X};\boldsymbol{\theta}^{\ast})\}\\
 & =\int\frac{h(Y,\boldsymbol{X};\boldsymbol{\theta}^{\ast})P(R=1)}{\pi(Y,\boldsymbol{X})}\left\{ \frac{\pi(Y,\boldsymbol{X})}{P(R=1)}f(Y,\boldsymbol{X},\boldsymbol{Z})\right\} dYd\boldsymbol{X}d\boldsymbol{Z}\\
 & =P(R=1)E\left\{ \frac{h(Y,\boldsymbol{X};\boldsymbol{\theta}^{\ast})}{\pi(Y,\boldsymbol{X})}\Big|R=1\right\} \\
 & =P(R=1)E\left[E\left\{ \frac{h(Y,\boldsymbol{X};\boldsymbol{\theta}^{\ast})}{\pi(Y,\boldsymbol{X})}\Big|\boldsymbol{X},\boldsymbol{Z},R=1 \right\} \Big|R=1\right]\\
 & =P(R=1)E\{\boldsymbol{u}(\boldsymbol{X},\boldsymbol{Z};\boldsymbol{\beta}_{0},\boldsymbol{\theta}^{\ast})|R=1\},
\end{align*}
where 
\[
\boldsymbol{u}(\boldsymbol{X},\boldsymbol{Z};\boldsymbol{\beta},\boldsymbol{\theta})=\int\frac{h(Y,\boldsymbol{X};\boldsymbol{\theta})}{\pi(Y,\boldsymbol{X})}f_{c}(Y|\boldsymbol{X},\boldsymbol{Z},R=1;\boldsymbol{\beta})dY,
\]
$f_{c}(Y|\boldsymbol{X},\boldsymbol{Z},R=1;\boldsymbol{\beta})$
is given in (\ref{condden}) and $\boldsymbol{\beta}_{0}$ is the
true value for $\boldsymbol{\beta}$ such that $f(Y|\boldsymbol{X},\boldsymbol{Z};\boldsymbol{\beta}_{0})=f(Y|\boldsymbol{X},\boldsymbol{Z})$.
Therefore, the information in (\ref{phase1_infor}) can be transformed
into 
\begin{equation}
E_{(\boldsymbol{X},\boldsymbol{Z}|R=1)}\{\boldsymbol{u}(\boldsymbol{X},\boldsymbol{Z};\boldsymbol{\beta}_{0},\boldsymbol{\theta}^{\ast})|R=1\}={\bf 0},\label{ufunc1}
\end{equation}
where to make it explicit we write $E_{(\boldsymbol{X},\boldsymbol{Z}|R=1)}(\cdot|R=1)$
as the expectation taken under the conditional covariate distribution
$F(\boldsymbol{X},\boldsymbol{Z}|R=1)$. 

The moment condition (\ref{ufunc1}) contains information about 
the parameter of interest $\boldsymbol{\beta}$ but depends on the conditional distribution $F(\boldsymbol{X},\boldsymbol{Z}|R=1)$ of the covariates. 
To incorporate this information about $\boldsymbol{\beta}$,
we consider an empirical distribution for $F(\boldsymbol{X},\boldsymbol{Z}|R=1)$,
$p_{i}=dF(\boldsymbol{X}_{i},\boldsymbol{Z}_{i}|R_{i}=1)$, $i=1,\ldots,m$.
Here $m$ is the sample size for Phase 2 data and, without loss of
generality, we index the Phase 2 subjects as $i=1,\ldots,m$. 

We can then define an estimator $\widehat{\boldsymbol{\beta}}_{\text{\ensuremath{\pi\theta^{\ast}}-1}}$
through 
\begin{align}
 & \max_{\boldsymbol{\beta},p_{1},...,p_{m}}\prod_{i=1}^{m}f_{c}(Y_{i}|\boldsymbol{X}_{i},\boldsymbol{Z}_{i},R_{i}=1;\boldsymbol{\beta})p_{i}\ \mbox{ subject to }\nonumber \\
 & p_{i}>0,\ \sum_{i=1}^{m}p_{i}=1,\ \mbox{and}\ \sum_{i=1}^{m}p_{i}\boldsymbol{u}(\boldsymbol{X}_{i},\boldsymbol{Z}_{i};\boldsymbol{\beta},\boldsymbol{\theta}^{\ast})={\bf 0}.\label{EL0-1}
\end{align}
The formulation in (\ref{EL0-1}) maximizes the likelihood function
corresponding to the conditional distribution $f(Y,\boldsymbol{X},\boldsymbol{Z}|R=1)=f(Y|\boldsymbol{X},\boldsymbol{Z},R=1)f(\boldsymbol{X},\boldsymbol{Z}|R=1)$,
where $f(Y|\boldsymbol{X},\boldsymbol{Z},R=1)$ is modeled parametrically and $f(\boldsymbol{X},\boldsymbol{Z}|R=1)$
is modeled nonparametrically subject to certain constraints. Such
a formulation is similar to that in \cite{qin2000combining,chatterjee2016constrained,han2019empirical}. 

In the Appendix we show that, as $n\to\infty$, $\sqrt{n}(\widehat{\boldsymbol{\beta}}_{\text{\ensuremath{\pi\theta^{\ast}}-1}}-\boldsymbol{\beta}_0)$ has an asymptotic normal distribution with mean zero and variance  $\boldsymbol{\Sigma}_{\pi\theta^{\ast}}=(\boldsymbol{S}+\boldsymbol{J}\boldsymbol{\Omega}^{-1}\boldsymbol{J}^{T})^{-1}$, where
$\boldsymbol{S}=E\{R\boldsymbol{s}_{c,\beta}(\boldsymbol{\beta}_{0})\boldsymbol{s}_{c,\beta}(\boldsymbol{\beta}_{0})^{T}\}$
with $\boldsymbol{s}_{c,\beta}(\boldsymbol{\beta})=\partial\log\{f_{c}(Y|\boldsymbol{X},\boldsymbol{Z},R=1;\boldsymbol{\beta})\}/\partial\boldsymbol{\beta}$
being the score function for $f_{c}(Y|\boldsymbol{X},\boldsymbol{Z},R=1;\boldsymbol{\beta})$
given in (\ref{condden}), $\boldsymbol{J}=E\{R\partial\boldsymbol{u}(\boldsymbol{\beta}_{0},\boldsymbol{\theta}^{*})/\partial\boldsymbol{\beta}^{T}\}$,
and $\boldsymbol{\Omega}=E\{R\boldsymbol{u}(\boldsymbol{\beta}_{0},\boldsymbol{\theta}^{*})\boldsymbol{u}(\boldsymbol{\beta}_{0},\boldsymbol{\theta}^{*})^{T}\}$.
The CML estimator $\widehat{\boldsymbol{\beta}}_{\text{CML}}$ that
maximizes (\ref{CML lik}) has the asymptotic distribution $\sqrt{n}(\widehat{\boldsymbol{\beta}}_{\text{CML}}-\boldsymbol{\beta}_{0})\xrightarrow{d}N({\bf 0},\boldsymbol{S}^{-1})$
as $n\to\infty$. It is easy to see that $\boldsymbol{\Sigma}_{\pi\theta^{\ast}}$
is smaller than $\boldsymbol{S}^{-1}$ in the positive-definite sense,
and thus $\widehat{\boldsymbol{\beta}}_{\text{\ensuremath{\pi\theta^{\ast}}-1}}$
is guaranteed to be more efficient than $\widehat{\boldsymbol{\beta}}_{\text{CML}}$
due to the incorporation of the Phase 1 information. 

Following \cite{han2016comment}, a formulation alternative to (\ref{EL0-1})
that is in line with \cite{qin1994empirical} can be considered. Specifically,
consider another estimator $\widehat{\boldsymbol{\beta}}_{\text{\ensuremath{\pi\theta^{\ast}}-2}}$
defined through 
\begin{align}
 \max_{\boldsymbol{\beta},p_{1},...,p_{m}}\prod_{i=1}^{m}p_{i}\ \mbox{ subject to }p_{i}>0,\ \sum_{i=1}^{m}p_{i}=1,
\mbox{and}\  & \sum_{i=1}^{m}p_{i}\left(\begin{array}{c}
\boldsymbol{s}_{c,\beta}(Y_{i},\boldsymbol{X}_{i},\boldsymbol{Z}_{i};\boldsymbol{\beta})\\
\boldsymbol{u}(\boldsymbol{X}_{i},\boldsymbol{Z}_{i};\boldsymbol{\beta},\boldsymbol{\theta}^{*})
\end{array}\right)={\bf 0},\label{EL0-2}
\end{align}
where, with a little abuse of notation, $p_{i}=dF(Y_{i},\boldsymbol{X}_{i},\boldsymbol{Z}_{i}|R_{i}=1)$,
$i=1,\ldots,m$, are an empirical distribution for $F(Y,\boldsymbol{X},\boldsymbol{Z}|R=1)$.
Using the result in \cite{qin1994empirical}, it is easy to show that $\widehat{\boldsymbol{\beta}}_{\text{\ensuremath{\pi\theta^{\ast}}-2}}$
is asymptotically equivalent to $\widehat{\boldsymbol{\beta}}_{\text{\ensuremath{\pi\theta^{\ast}}-1}}$
in the sense that $\sqrt{n}(\widehat{\boldsymbol{\beta}}_{\text{\ensuremath{\pi\theta^{\ast}}-2}}-\boldsymbol{\beta}_{0})\xrightarrow{d}N({\bf 0},\boldsymbol{\Sigma}_{\pi\theta^{\ast}})$
as $n\to\infty$ with the same asymptotic variance $\boldsymbol{\Sigma}_{\pi\theta^{\ast}}$.

In both (\ref{EL0-1}) and (\ref{EL0-2}) $\boldsymbol{\theta}^{*}$
is used instead of $\widehat{\boldsymbol{\theta}}$, and thus the
uncertainty associated with $\widehat{\boldsymbol{\theta}}$ resulting
from Phase 1 data is not accounted for by the two estimators $\widehat{\boldsymbol{\beta}}_{\text{\ensuremath{\pi\theta^{\ast}}-1}}$
and $\widehat{\boldsymbol{\beta}}_{\text{\ensuremath{\pi\theta^{\ast}}-2}}$.
This is less of a concern when Phase 1 sample size is large and only
a small fraction of the Phase 1 subjects are selected to enter Phase
2. However, when the Phase 1 sample size is not very large the uncertainty
associated with $\widehat{\boldsymbol{\theta}}$ needs to be properly
accounted for. 

Specifically, let $\widehat{\boldsymbol{\beta}}_{\text{\ensuremath{\pi\widehat{\theta}}-1}}$
and $\widehat{\boldsymbol{\beta}}_{\text{\ensuremath{\pi\widehat{\theta}}-2}}$
denote the estimators by replacing the $\boldsymbol{\theta}^{*}$
in (\ref{EL0-1}) and (\ref{EL0-2}) with $\widehat{\boldsymbol{\theta}}$,
respectively. In the Appendix we show that both $\sqrt{n}(\widehat{\boldsymbol{\beta}}_{\text{\ensuremath{\pi\widehat{\theta}}-1}}-\boldsymbol{\beta}_{0})$
and $\sqrt{n}(\widehat{\boldsymbol{\beta}}_{\text{\ensuremath{\pi\widehat{\theta}}-2}}-\boldsymbol{\beta}_{0})$
converge in distribution to $N({\bf 0},\boldsymbol{\Sigma}_{\pi\widehat{\theta}})$
as $n\to\infty$, and thus $\widehat{\boldsymbol{\beta}}_{\text{\ensuremath{\pi\widehat{\theta}}-1}}$
and $\widehat{\boldsymbol{\beta}}_{\text{\ensuremath{\pi\widehat{\theta}}-2}}$
are also asymptotically equivalent. Here 
\begin{align*}
\boldsymbol{\Sigma}_{\pi\widehat{\theta}}&=(\boldsymbol{S}+\boldsymbol{J}\boldsymbol{\Omega}^{-1}\boldsymbol{J}^{T})^{-1}\left\{ \boldsymbol{S}+\boldsymbol{J}\boldsymbol{\Omega}^{-1}\boldsymbol{U}+\boldsymbol{U}^{T}\boldsymbol{\Omega}^{-1}\boldsymbol{J}^{T}+\boldsymbol{J}\boldsymbol{\Omega}^{-1}(\boldsymbol{\Omega}-\boldsymbol{V}-\boldsymbol{V}^{T}+\boldsymbol{W})\boldsymbol{\Omega}^{-1}\boldsymbol{J}^{T}\right\}\\
&(\boldsymbol{S}+\boldsymbol{J}\boldsymbol{\Omega}^{-1}\boldsymbol{J}^{T})^{-1},
\end{align*}
$\boldsymbol{J}$ and $\boldsymbol{\Omega}$ are the same as before,
$\boldsymbol{U}=E\{R\boldsymbol{s}_{c,\beta}(\boldsymbol{\beta}_{0})\boldsymbol{h}(\boldsymbol{\theta}^{\ast})^{T}\}$,
$\boldsymbol{V}=E\{R\boldsymbol{u}(\boldsymbol{\beta}_{0},\boldsymbol{\theta}^{\ast})\boldsymbol{h}(\boldsymbol{\theta}^{\ast})^{T}\}$,
and $\boldsymbol{W}=E\{\boldsymbol{h}(\boldsymbol{\theta}^{\ast})\boldsymbol{h}(\boldsymbol{\theta}^{\ast})^{T}\}$.
There is no direct comparison between $\boldsymbol{\Sigma}_{\pi\widehat{\theta}}$
and $\boldsymbol{\Sigma}_{\pi\theta^{\ast}}$ or $\boldsymbol{S}^{-1}$
due to its complex form. In the next section we will construct estimators
whose asymptotic variances are smaller than both $\boldsymbol{\Sigma}_{\pi\widehat{\theta}}$
and $\boldsymbol{S}^{-1}$. 

\subsection{Modeling the selection probability}% $\pi(Y,\boldsymbol{X})$}

For two-phase ODS, the selection probability $\pi(Y,\boldsymbol{X})$
is known by design and the estimators in the previous subsection all
use the known $\pi(Y,\boldsymbol{X})$. For the CML method some authors
have found that modeling $\pi(Y,\boldsymbol{X})$ and using the estimated
value instead can further improve estimation efficiency (e.g., \cite{scott1997fitting,lawless1999semiparametric,scott2011fitting}). This observation has also
been made in the missing data literature for the IPW/AIPW method (e.g.,
\cite{robins1994estimation}). Following these observations, we consider postulating
a parametric model $\pi(Y,\boldsymbol{X};\boldsymbol{\alpha})$ for
$\pi(Y,\boldsymbol{X})$ and construct estimators that further improve
over the previous ones. Here the parameter $\boldsymbol{\alpha}$
has true value $\boldsymbol{\alpha}_{0}$ such that $\pi(Y,\boldsymbol{X};\boldsymbol{\alpha}_{0})=\pi(Y,\boldsymbol{X})$.
Although the $\boldsymbol{\alpha}_{0}$ is known by design, in this
subsection we estimate it using the collected data. The most straightforward
way is to first estimate $\boldsymbol{\alpha}_{0}$ by $\widehat{\boldsymbol{\alpha}}_{\text{MLE}}$
that maximizes the likelihood 
\begin{equation}
\prod_{i=1}^{n}\pi(Y_{i},\boldsymbol{X}_{i};\boldsymbol{\alpha})^{R_{i}}\{1-\pi(Y_{i},\boldsymbol{X}_{i};\boldsymbol{\alpha})\}^{1-R_{i}}\label{pi lik}
\end{equation}
and then replace the $\pi(Y,\boldsymbol{X})$ by $\pi(Y,\boldsymbol{X};\widehat{\boldsymbol{\alpha}}_{\text{MLE}})$
in all relevant derivations leading to (\ref{EL0-1}) and (\ref{EL0-2}).
This is similar to the construction of $\widehat{\boldsymbol{\beta}}_{\text{\ensuremath{\pi\widehat{\theta}}-1}}$
and $\widehat{\boldsymbol{\beta}}_{\text{\ensuremath{\pi\widehat{\theta}}-2}}$
in Section 2.1 by plugging in $\widehat{\boldsymbol{\theta}}$ that
is estimated separately beforehand. 

With $\pi(Y,\boldsymbol{X})$ modeled by $\pi(Y,\boldsymbol{X};\boldsymbol{\alpha})$,
the $f_{c}(Y|\boldsymbol{X},\boldsymbol{Z},R=1;\boldsymbol{\beta})$
in (\ref{condden}) should also depend on $\boldsymbol{\alpha}$,
and we rewrite it as $f_{c}(Y|\boldsymbol{X},\boldsymbol{Z},R=1;\boldsymbol{\beta},\boldsymbol{\alpha})$
to make this dependence explicit. Correspondingly, the score function
for $\boldsymbol{\beta}$ is $\boldsymbol{s}_{c,\beta}(\boldsymbol{\beta},\boldsymbol{\alpha})=\partial\log\{f_{c}(Y|\boldsymbol{X},\boldsymbol{Z},R=1;\boldsymbol{\beta},\boldsymbol{\alpha})\}/\partial\boldsymbol{\beta}$,
previously denoted as $\boldsymbol{s}_{c,\beta}(\boldsymbol{\beta})$
when the known $\pi(Y,\boldsymbol{X})$ was used. Similarly, the $\boldsymbol{u}(\boldsymbol{X},\boldsymbol{Z};\boldsymbol{\beta},\boldsymbol{\theta})$
in (\ref{ufunc1}) becomes $\boldsymbol{u}(\boldsymbol{X},\boldsymbol{Z};\boldsymbol{\beta},\boldsymbol{\alpha},\boldsymbol{\theta})$.
From the theory of estimating functions and semiparametric efficiency
(e.g., \cite{tsiatis2007semiparametric}), estimating nuisance parameters separately beforehand
and then plugging them in does not lead to the maximum efficiency.
The maximum efficiency is achieved when the parameter of interest
$\boldsymbol{\beta}$ is estimated jointly with the nuisance parameters
$\boldsymbol{\alpha}$ and $\boldsymbol{\theta}$, with the score
function corresponding to the likelihood in (\ref{pi lik}) and the
score function $\boldsymbol{h}(Y,\boldsymbol{X};\boldsymbol{\theta})$
for $f(Y|\boldsymbol{X};\boldsymbol{\theta})$ added as additional
estimating functions for $\boldsymbol{\alpha}$ and $\boldsymbol{\theta}$,
in addition to $\boldsymbol{s}_{c,\beta}(\boldsymbol{\beta},\boldsymbol{\alpha})$
and $\boldsymbol{u}(\boldsymbol{X},\boldsymbol{Z};\boldsymbol{\beta},\boldsymbol{\alpha},\boldsymbol{\theta})$. 

Specifically, let $\boldsymbol{s}_{\alpha}(Y,\boldsymbol{X},R;\boldsymbol{\alpha})$ %=\frac{R-\pi(Y,\boldsymbol{X};\boldsymbol{\alpha})}{\pi(Y,\boldsymbol{X};\boldsymbol{\alpha})\{1-\pi(Y,\boldsymbol{X};\boldsymbol{\alpha})\}}\frac{\partial\pi(Y,\boldsymbol{X};\boldsymbol{\alpha})}{\partial\boldsymbol{\alpha}}\]
 denote the score function for $\boldsymbol{\alpha}$ corresponding
to (\ref{pi lik}). We can define an estimator $\widehat{\boldsymbol{\beta}}_{\text{\ensuremath{\widehat{\pi}\widehat{\theta}}-3}}$
through 
\begin{align}
 & \max_{\boldsymbol{\beta},\boldsymbol{\alpha},\boldsymbol{\theta},q_{1},...,q_{n}}\prod_{i=1}^{n}q_{i}\ \mbox{ subject to }\nonumber \\
 & q_{i}\ge0,\ \sum_{i=1}^{n}q_{i}=1,\ \mbox{and}\ \sum_{i=1}^{n}q_{i}\left(\begin{array}{l}
R_{i}\boldsymbol{s}_{c,\beta}(Y_{i},\boldsymbol{X}_{i},\boldsymbol{Z}_{i};\boldsymbol{\beta},\boldsymbol{\alpha})\\
R_{i}\boldsymbol{u}(\boldsymbol{X}_{i},\boldsymbol{Z}_{i};\boldsymbol{\beta},\boldsymbol{\alpha},\boldsymbol{\theta})\\
\boldsymbol{s}_{\alpha}(Y_{i},\boldsymbol{X}_{i},R_{i};\boldsymbol{\alpha})\\
\boldsymbol{h}(Y_{i},\boldsymbol{X}_{i};\boldsymbol{\theta})
\end{array}\right)={\bf 0},\label{EL3}
\end{align}
where $q_{i}=dF(Y_{i},\boldsymbol{X}_{i},\boldsymbol{Z}_{i},R_{i})$,
$i=1,\ldots,n$. The $\widehat{\boldsymbol{\beta}}_{\text{\ensuremath{\widehat{\pi}\widehat{\theta}}-3}}$
is based on a joint maximization with respect to $\boldsymbol{\beta}$,
$\boldsymbol{\alpha}$ and $\boldsymbol{\theta}$ and makes use of
information available in the form of estimating functions. From the
theory of estimating functions and semiparametric efficiency (e.g.,\cite{tsiatis2007semiparametric}), $\widehat{\boldsymbol{\beta}}_{\text{\ensuremath{\widehat{\pi}\widehat{\theta}}-3}}$
is more efficient than all previous estimators. 

There are ways to make a further efficiency improvement over $\widehat{\boldsymbol{\beta}}_{\text{\ensuremath{\widehat{\pi}\widehat{\theta}}-3}}$.
As \cite{scott2011fitting} pointed out, the score function component
$\boldsymbol{s}_{c,\alpha}(\boldsymbol{\beta},\boldsymbol{\alpha})=\partial\log\{f_{c}(Y|\boldsymbol{X},\boldsymbol{Z},R=1;\boldsymbol{\beta},\boldsymbol{\alpha})\}/\partial\boldsymbol{\alpha}$
also contains information about both $\boldsymbol{\beta}$ and $\boldsymbol{\alpha}$,
similar to $\boldsymbol{s}_{c,\beta}(\boldsymbol{\beta},\boldsymbol{\alpha})$.
Therefore, $\boldsymbol{s}_{c,\alpha}(\boldsymbol{\beta},\boldsymbol{\alpha})$
should be similarly accounted for as an estimating function. Specifically,
we can define an estimator $\widehat{\boldsymbol{\beta}}_{\text{\ensuremath{\widehat{\pi}\widehat{\theta}}-4}}$
by replacing the third set of constraints in (\ref{EL3}) with 
\begin{equation}
\sum_{i=1}^{n}q_{i}\left(\begin{array}{l}
R_{i}\boldsymbol{s}_{c,\beta}(Y_{i},\boldsymbol{X}_{i},\boldsymbol{Z}_{i};\boldsymbol{\beta},\boldsymbol{\alpha})\\
R_{i}\boldsymbol{u}(\boldsymbol{X}_{i},\boldsymbol{Z}_{i};\boldsymbol{\beta},\boldsymbol{\alpha},\boldsymbol{\theta})\\
R_{i}\boldsymbol{s}_{c,\alpha}(Y_{i},\boldsymbol{X}_{i},\boldsymbol{Z}_{i};\boldsymbol{\beta},\boldsymbol{\alpha})\\
\boldsymbol{s}_{\alpha}(Y_{i},\boldsymbol{X}_{i},R_{i};\boldsymbol{\alpha})\\
\boldsymbol{h}(Y_{i},\boldsymbol{X}_{i};\boldsymbol{\theta})
\end{array}\right)={\bf 0}.
\label{EL4}
\end{equation}
The $\widehat{\boldsymbol{\beta}}_{\text{\ensuremath{\widehat{\pi}\widehat{\theta}}-4}}$
makes use of every piece of information available in the form of an
estimating function, and thus has the maximum efficiency we are able
to achieve in estimating $\boldsymbol{\beta}$ under the model assumptions we have made. 

A potential issue with $\widehat{\boldsymbol{\beta}}_{\text{\ensuremath{\widehat{\pi}\widehat{\theta}}-4}}$
is that in some special cases, especially when the outcome $Y$ is
binary, there may exist multicollinearity among the components of
$\boldsymbol{s}_{c,\alpha}(Y,\boldsymbol{X},\boldsymbol{Z};\boldsymbol{\beta},\boldsymbol{\alpha})$
\cite{che2020empirical}. The estimator proposed in \cite{scott2011fitting}
by solving 
\begin{equation}
\frac{1}{n}\sum_{i=1}^{n}\left(\begin{array}{l}
R_{i}\boldsymbol{s}_{c,\beta}(Y_{i},\boldsymbol{X}_{i},\boldsymbol{Z}_{i};\boldsymbol{\beta},\boldsymbol{\alpha})\\
\boldsymbol{s}_{\alpha}(Y_{i},\boldsymbol{X}_{i},R_{i};\boldsymbol{\alpha})-R_{i}\boldsymbol{s}_{c,\alpha}(Y_{i},\boldsymbol{X}_{i},\boldsymbol{Z}_{i};\boldsymbol{\beta},\boldsymbol{\alpha})
\end{array}\right)={\bf 0}\label{SW}
\end{equation}
avoids any potential multicollinearity. We follow their idea and consider
another estimator $\widehat{\boldsymbol{\beta}}_{\text{\ensuremath{\widehat{\pi}\widehat{\theta}}-5}}$
defined through 

\begin{align}
 & \max_{\boldsymbol{\beta},\boldsymbol{\alpha},\boldsymbol{\theta},q_{1},...,q_{n}}\prod_{i=1}^{n}q_{i}\ \mbox{ subject to }\nonumber \\
 & q_{i}\ge0,\ \sum_{i=1}^{n}q_{i}=1,\ \mbox{and}\ \sum_{i=1}^{n}q_{i}\left(\begin{array}{l}
R_{i}\boldsymbol{s}_{c,\beta}(Y_{i},\boldsymbol{X}_{i},\boldsymbol{Z}_{i};\boldsymbol{\beta},\boldsymbol{\alpha})\\
R_{i}\boldsymbol{u}(\boldsymbol{X}_{i},\boldsymbol{Z}_{i};\boldsymbol{\beta},\boldsymbol{\alpha},\boldsymbol{\theta})\\
\boldsymbol{s}_{\alpha}(Y_{i},\boldsymbol{X}_{i},R_{i};\boldsymbol{\alpha})-R_{i}\boldsymbol{s}_{c,\alpha}(Y_{i},\boldsymbol{X}_{i},\boldsymbol{Z}_{i};\boldsymbol{\beta},\boldsymbol{\alpha})\\
\boldsymbol{h}(Y_{i},\boldsymbol{X}_{i};\boldsymbol{\theta})
\end{array}\right)={\bf 0}.\label{EL5}
\end{align}
In the absence of multicollinearity, $\widehat{\boldsymbol{\beta}}_{\text{\ensuremath{\widehat{\pi}\widehat{\theta}}-4}}$
has a smaller asymptotic variance compared to $\widehat{\boldsymbol{\beta}}_{\text{\ensuremath{\widehat{\pi}\widehat{\theta}}-5}}$.
However, the formulation in (\ref{EL5}) avoids potential multicollinearity
and the resulting $\widehat{\boldsymbol{\beta}}_{\text{\ensuremath{\widehat{\pi}\widehat{\theta}}-5}}$
still has a high efficiency. Our recommendation for use is $\widehat{\boldsymbol{\beta}}_{\text{\ensuremath{\widehat{\pi}\widehat{\theta}}-5}}$.
An estimate of the asymptotic variance of $\widehat{\boldsymbol{\beta}}_{\text{\ensuremath{\widehat{\pi}\widehat{\theta}}-5}}$
can be obtained by directly applying the asymptotic variance formula
in \cite{qin1994empirical} and then taking the block matrix corresponding
to $\boldsymbol{\beta}$. Specifically, denoting all the estimating function as $\boldsymbol{U}(\boldsymbol{\eta})$ where $\boldsymbol{\eta}=(\boldsymbol{\beta}^T,\boldsymbol{\alpha}^T,\boldsymbol{\theta}^T)^T$, the asymptotic variance can be estimated by 
\[\left[E_n\left\{\frac{\partial \boldsymbol{U}(\widehat{\boldsymbol{\eta}})}{\partial\boldsymbol{\eta}^T}\right\}E_n\left\{\boldsymbol{U}(\widehat{\boldsymbol{\eta}})\boldsymbol{U}(\widehat{\boldsymbol{\eta}})^T\right\}^{-1}E_n\left\{\frac{\partial \boldsymbol{U}(\widehat{\boldsymbol{\eta}})}{\partial\boldsymbol{\eta}}\right\}\right]^{-1},\]
where $E_n$ stands for the empirical expectation that can be evaluated by taking the sample mean, and the differentiation can be evaluated numerically. The efficiencies of $\widehat{\boldsymbol{\beta}}_{\text{\ensuremath{\widehat{\pi}\widehat{\theta}}-4}}$ and $\widehat{\boldsymbol{\beta}}_{\text{\ensuremath{\widehat{\pi}\widehat{\theta}}-5}}$ are guaranteed to improve upon $\widehat{\boldsymbol{\beta}}_{\text{\ensuremath{\widehat{\pi}\widehat{\theta}}-2}}$ by Corollary 1 in \cite{qin1994empirical}. Expressions for the asymptotic variances in terms of block matrices similar to $\boldsymbol{\Sigma}_{\pi\widehat{\theta}}$ involve a large number of submatrices and we do not give an expression here.

\section{The Method With Zero Selection Probabilities}

We now consider the case where some Phase 1 subjects are excluded
from entering Phase 2 by design or, in other words, have a zero probability
of entering Phase 2. Let $\mathcal{D}$ denote the set of values in
the range of $Y$ that lead to positive selection probabilities to
enter Phase 2. For example, for a continuous outcome $Y$ that can
take any real values, suppose the two-phase ODS only samples subjects
with outcome values smaller than $-2$ or larger than $2$ to enter
Phase 2, then $\mathcal{D}=\{y:y<-2\text{ or }y>2\}$. Let $S=I(Y\in\mathcal{D})$
denote the indicator for having a positive probability of being selected
to Phase 2. The observed data are now a random sample $(Y_{i},\boldsymbol{X}_{i},S_{i},S_{i}R_{i},S_{i}R_{i}\boldsymbol{Z}_{i})$,
$i=1,\ldots,n$. In this case, let $\pi(Y,\boldsymbol{X})=P(R=1|Y,\boldsymbol{X},S=1)>0$
denote the positive selection probability for subjects with $S=1$,
then we have $P(R=1|Y,\boldsymbol{X},S)=S\pi(Y,\boldsymbol{X})$.
Based on the observed data, the CML estimator $\widehat{\boldsymbol{\beta}}_{\text{CML}}$
now is the maximizer of the conditional likelihood 
\[
\prod_{i=1}^{n}f_{cc}(Y_{i}|\boldsymbol{X}_{i},\boldsymbol{Z}_{i},S_{i}=1,R_{i}=1;\boldsymbol{\beta})^{S_{i}R_{i}},
\]
where 
\begin{align*}
f_{cc}(Y|\boldsymbol{X},\boldsymbol{Z},S=1,R=1;\boldsymbol{\beta})=\frac{f(Y|\boldsymbol{X},\boldsymbol{Z};\boldsymbol{\beta})\pi(Y,\boldsymbol{X})I(Y\in\mathcal{D})}{\int f(Y|\boldsymbol{X},\boldsymbol{Z};\boldsymbol{\beta})\pi(Y,\boldsymbol{X})I(Y\in\mathcal{D})dY}.
\end{align*}
It is worth pointing out that, for case-only studies in the literature
where $Y$ is binary and only the cases are selected (e.g., Piegorsch
et al. 1994), the conditional density $f_{cc}(Y|\boldsymbol{X},\boldsymbol{Z},S=1,R=1;\boldsymbol{\beta})$
degenerates. Therefore the CML method, and hence our proposed method,
does not work for case-only studies. 

In the presence of zero selection probabilities, to incorporate the
Phase 1 information summarized as (\ref{phase1_infor}) into the estimation
of $\boldsymbol{\beta}$ and have an efficiency improvement over the
CML estimator, we now need to assume the working model $f(Y|\boldsymbol{X};\boldsymbol{\theta})$
for $f(Y|\boldsymbol{X})$ is correctly specified such that $f(Y|\boldsymbol{X};\boldsymbol{\theta}^{\ast})=f(Y|\boldsymbol{X})$.
To make this assumption explicit in our notation, we will switch from
$\boldsymbol{\theta}^{\ast}$ to $\boldsymbol{\theta}_{0}$ to denote
the true value of $\boldsymbol{\theta}$, which is still the probability
limit of $\widehat{\boldsymbol{\theta}}$.

Define 
\begin{align*}
\boldsymbol{h}^{*}(\boldsymbol{X};\boldsymbol{\theta}_{0}) & =E\{\boldsymbol{h}(Y,\boldsymbol{X};\boldsymbol{\theta}_{0})|\boldsymbol{X},S=1\}\\
& =\int\boldsymbol{h}(Y,\boldsymbol{X};\boldsymbol{\theta}_{0})\frac{P(S=1|Y,\boldsymbol{X})f(Y|\boldsymbol{X})}{P(S=1|\boldsymbol{X})}dY\\
& =\frac{\int\boldsymbol{h}(Y,\boldsymbol{X};\boldsymbol{\theta}_{0})f(Y|\boldsymbol{X};\boldsymbol{\theta}_{0})I(Y\in\mathcal{D})dY}{\int f(Y|\boldsymbol{X};\boldsymbol{\theta}_{0})I(Y\in\mathcal{D})dY},
\end{align*}
then we must have $E\{\boldsymbol{h}(Y,\boldsymbol{X};\boldsymbol{\theta}_{0})-\boldsymbol{h}^{*}(\boldsymbol{X};\boldsymbol{\theta}_{0})|\boldsymbol{X},S=1\}=\boldsymbol{0}$,
which implies that $E\{\boldsymbol{h}(Y,\boldsymbol{X};\boldsymbol{\theta}_{0})-\boldsymbol{h}^{*}(\boldsymbol{X};\boldsymbol{\theta}_{0})|S=1\}=\boldsymbol{0}$.
Therefore, we have 
\begin{align*}
{\bf 0} & =E\{\boldsymbol{h}(Y,\boldsymbol{X};\boldsymbol{\theta}_{0})-\boldsymbol{h}^{*}(\boldsymbol{X};\boldsymbol{\theta}_{0})|S=1\}\\
& =\int P(R=1|S=1)\frac{\boldsymbol{h}(Y,\boldsymbol{X};\boldsymbol{\theta}_{0})-\boldsymbol{h}^{*}(\boldsymbol{X};\boldsymbol{\theta}_{0})}{\pi(Y,\boldsymbol{X})}\left\{ \frac{P(R=1|Y,\boldsymbol{X},\boldsymbol{Z},S=1)f(Y,\boldsymbol{X},\boldsymbol{Z}|S=1)}{P(R=1|S=1)}\right\} \\&dYd\boldsymbol{X}d\boldsymbol{Z}\\
& =P(R=1|S=1)E\left\{ \frac{\boldsymbol{h}(Y,\boldsymbol{X};\boldsymbol{\theta}_{0})-\boldsymbol{h}^{*}(\boldsymbol{X};\boldsymbol{\theta}_{0})}{\pi(Y,\boldsymbol{X})}\Big|S=1,R=1\right\} \\
& =P(R=1|S=1)E\left[E\left\{ \frac{\boldsymbol{h}(Y,\boldsymbol{X};\boldsymbol{\theta}_{0})-\boldsymbol{h}^{*}(\boldsymbol{X};\boldsymbol{\theta}_{0})}{\pi(Y,\boldsymbol{X})}\Big|\boldsymbol{X},\boldsymbol{Z},S=1,R=1\right\} \Big|S=1,R=1\right]\\
& =P(R=1|S=1)E\{\boldsymbol{v}(\boldsymbol{X},\boldsymbol{Z};\boldsymbol{\beta}_{0},\boldsymbol{\theta}_{0})|S=1,R=1\},
\end{align*}
\begin{align*}& \mbox{where}\ \boldsymbol{v}(\boldsymbol{X},\boldsymbol{Z};\boldsymbol{\beta},\boldsymbol{\theta}) =E\left\{ \frac{\boldsymbol{h}(Y,\boldsymbol{X};\boldsymbol{\theta})-\boldsymbol{h}^{*}(\boldsymbol{X};\boldsymbol{\theta})}{\pi(Y,\boldsymbol{X})}\Big|\boldsymbol{X},\boldsymbol{Z},S=1,R=1\right\} \\
 & =\int\frac{\boldsymbol{h}(Y,\boldsymbol{X};\boldsymbol{\theta})-\boldsymbol{h}^{*}(\boldsymbol{X};\boldsymbol{\theta})}{\pi(Y,\boldsymbol{X})}f_{cc}(Y|\boldsymbol{X},\boldsymbol{Z},S=1,R=1;\boldsymbol{\beta})dY.
\end{align*}
In other words, the information in (\ref{phase1_infor}) is now summarized
as 
\begin{equation}
E_{(\boldsymbol{X},\boldsymbol{Z}|S=1,R=1)}\{\boldsymbol{v}(\boldsymbol{X},\boldsymbol{Z};\boldsymbol{\beta}_{0},\boldsymbol{\theta}_{0})|S=1,R=1\}={\bf 0},\label{ufunc2}
\end{equation}
where $E_{(\boldsymbol{X},\boldsymbol{Z}|S=1,R=1)}(\cdot|S=1,R=1)$
is the expectation taken under the conditional covariate distribution
$F(\boldsymbol{X},\boldsymbol{Z}|S=1,R=1)$. The moment condition
(\ref{ufunc2}) contains information about both the parameter of interest
$\boldsymbol{\beta}$ and the conditional distribution $F(\boldsymbol{X},\boldsymbol{Z}|S=1,R=1)$.
Therefore, estimators similar to those in Section 2 can be defined. 

Specifically, consider an empirical distribution for $F(\boldsymbol{X},\boldsymbol{Z}|S=1,R=1)$,
$p_{i}=dF(\boldsymbol{X}_{i},\boldsymbol{Z}_{i}|S_{i}=1,R_{i}=1)$,
$i=1,\ldots,m$. Then estimator $\widehat{\boldsymbol{\beta}}_{\text{\ensuremath{\pi\theta_{0}}-1}}$
can be redefined through 
\begin{align}
 & \max_{\boldsymbol{\beta},p_{1},...,p_{m}}\prod_{i=1}^{m}f_{cc}(Y_{i}|\boldsymbol{X}_{i},\boldsymbol{Z}_{i},S_{i}=1,R_{i}=1;\boldsymbol{\beta})p_{i}\ \mbox{ subject to }\nonumber \\
 & p_{i}>0,\ \sum_{i=1}^{m}p_{i}=1,\ \mbox{and}\ \sum_{i=1}^{m}p_{i}\boldsymbol{v}(\boldsymbol{X}_{i},\boldsymbol{Z}_{i};\boldsymbol{\beta},\boldsymbol{\theta}_{0})={\bf 0}.\label{EL0-1-zero}
\end{align}
Alternatively, consider an empirical distribution for $F(Y,\boldsymbol{X},\boldsymbol{Z}|S=1,R=1)$,
$p_{i}=dF(Y_{i},\boldsymbol{X}_{i},\boldsymbol{Z}_{i}|S_{i}=1,R_{i}=1)$,
$i=1,\ldots,m$. Then $\widehat{\boldsymbol{\beta}}_{\text{\ensuremath{\pi\theta_{0}}-2}}$
can be redefined through 

\begin{align*}
 & \max_{\boldsymbol{\beta},p_{1},...,p_{m}}\prod_{i=1}^{m}p_{i}\ \mbox{ subject to }p_{i}>0,\ \sum_{i=1}^{m}p_{i}=1,\ \mbox{and}\  \sum_{i=1}^{m}p_{i}\left(\begin{array}{c}
\boldsymbol{s}_{cc,\beta}(Y_{i},\boldsymbol{X}_{i},\boldsymbol{Z}_{i};\boldsymbol{\beta})\\
\boldsymbol{v}(\boldsymbol{X}_{i},\boldsymbol{Z}_{i};\boldsymbol{\beta},\boldsymbol{\theta}_{0})
\end{array}\right)={\bf 0},
\end{align*}
where $\boldsymbol{s}_{cc,\beta}(\boldsymbol{\beta})=\partial\log\{f_{cc}(Y|\boldsymbol{X},\boldsymbol{Z},S=1,R=1;\boldsymbol{\beta})\}/\partial\boldsymbol{\beta}$
is the score function. It can be shown similarly to before that $\widehat{\boldsymbol{\beta}}_{\text{\ensuremath{\pi\theta_{0}}-1}}$
and $\widehat{\boldsymbol{\beta}}_{\text{\ensuremath{\pi\theta_{0}}-2}}$
are asymptotically equivalent with asymptotic variance $\widetilde{\boldsymbol{\Sigma}}_{\pi\theta_{0}}=(\widetilde{\boldsymbol{S}}+\widetilde{\boldsymbol{J}}\widetilde{\boldsymbol{\Omega}}^{-1}\widetilde{\boldsymbol{J}}^{T})^{-1}$,
where $\widetilde{\boldsymbol{S}}=E\{RS\boldsymbol{s}_{cc,\beta}(\boldsymbol{\beta}_{0})\boldsymbol{s}_{cc,\beta}(\boldsymbol{\beta}_{0})^{T}\}$,
$\widetilde{\boldsymbol{J}}=E\{RS\partial\boldsymbol{v}(\boldsymbol{\beta}_{0},\boldsymbol{\theta}_{0})/\partial\boldsymbol{\beta}^{T}\}$,
and $\widetilde{\boldsymbol{\Omega}}=E\{RS\boldsymbol{v}(\boldsymbol{\beta}_{0},\boldsymbol{\theta}_{0})\boldsymbol{v}(\boldsymbol{\beta}_{0},\boldsymbol{\theta}_{0})^{T}\}$.
Thus $\widetilde{\boldsymbol{\Sigma}}_{\pi\theta_{0}}$ is smaller
than the asymptotic variance $\widetilde{\boldsymbol{S}}$ of the
CML estimator $\widehat{\boldsymbol{\beta}}_{\text{CML}}$.

Estimators $\widehat{\boldsymbol{\beta}}_{\text{\ensuremath{\pi\widehat{\theta}}-1}}$
and $\widehat{\boldsymbol{\beta}}_{\text{\ensuremath{\pi\widehat{\theta}}-2}}$
can also be redefined by replacing $\boldsymbol{\theta}_{0}$ in the
constraints with $\widehat{\boldsymbol{\theta}}$. These two estimators
are again asymptotically equivalent, with asymptotic variance
\begin{align*}
\widetilde{\boldsymbol{\Sigma}}_{\pi\widehat{\theta}}=&(\widetilde{\boldsymbol{S}}+\widetilde{\boldsymbol{J}}\widetilde{\boldsymbol{\Omega}}^{-1}\widetilde{\boldsymbol{J}}^{T})^{-1}\{ \widetilde{\boldsymbol{S}}+\widetilde{\boldsymbol{J}}\widetilde{\boldsymbol{\Omega}}^{-1}\widetilde{\boldsymbol{H}}\boldsymbol{W}^{-1}\widetilde{\boldsymbol{U}}+\widetilde{\boldsymbol{U}}^{T}\boldsymbol{W}^{-1}\widetilde{\boldsymbol{H}}^{T}\widetilde{\boldsymbol{\Omega}}^{-1}\widetilde{\boldsymbol{J}}^{T}+\widetilde{\boldsymbol{J}}\widetilde{\boldsymbol{\Omega}}^{-1}(\widetilde{\boldsymbol{\Omega}}\\&-\widetilde{\boldsymbol{H}}\boldsymbol{W}^{-1}\widetilde{\boldsymbol{V}}
-\widetilde{\boldsymbol{V}}^{T}\boldsymbol{W}^{-1}\widetilde{\boldsymbol{H}}^{T}+\widetilde{\boldsymbol{H}}\boldsymbol{W}^{-1}\widetilde{\boldsymbol{H}}^{T})\widetilde{\boldsymbol{\Omega}}^{-1}\widetilde{\boldsymbol{J}}^{T}\}  (\widetilde{\boldsymbol{S}}+\widetilde{\boldsymbol{J}}\widetilde{\boldsymbol{\Omega}}^{-1}\widetilde{\boldsymbol{J}}^{T})^{-1},
\end{align*}
where $\widetilde{\boldsymbol{H}}=E\{RS\partial\boldsymbol{v}(\boldsymbol{\beta}_{0},\boldsymbol{\theta}_{0})/\partial\boldsymbol{\theta}^{T}\}$,
$\boldsymbol{W}=E\{\boldsymbol{h}(\boldsymbol{\theta}_{0})\boldsymbol{h}(\boldsymbol{\theta}_{0})^{T}\}$,
$\widetilde{\boldsymbol{U}}=E\{RS\boldsymbol{s}_{cc,\beta}(\boldsymbol{\beta}_{0},\boldsymbol{\theta}_{0})\boldsymbol{h}(\boldsymbol{\theta}_{0})^{T}\}$,
and $\widetilde{\boldsymbol{V}}=E\{RS\boldsymbol{v}(\boldsymbol{\beta}_{0},\boldsymbol{\theta}_{0})\boldsymbol{h}(\boldsymbol{\theta}_{0})^{T}\}$. 

When $\pi(Y,\boldsymbol{X})=P(R=1|Y,\boldsymbol{X},S=1)$ is modeled
parametrically as $\pi(Y,\boldsymbol{X};\boldsymbol{\alpha})$, estimators
with a further improved efficiency and similar to those in Section
2.2 can be constructed. In particular, let $q_{i}=dF(Y_{i},\boldsymbol{X}_{i},\boldsymbol{Z}_{i},S_{i},R_{i})$
denote an empirical distribution for $F(Y,\boldsymbol{X},\boldsymbol{Z},S,R)$,
$i=1,\ldots,n$, then $\widehat{\boldsymbol{\beta}}_{\text{\ensuremath{\widehat{\pi}\widehat{\theta}}-4}}$
can be redefined through 
\begin{align*}
 \max_{\boldsymbol{\beta},\boldsymbol{\alpha},\boldsymbol{\theta},q_{1},...,q_{n}}\prod_{i=1}^{n}q_{i}\ \mbox{ subject to }  q_{i}\ge0,\ \sum_{i=1}^{n}q_{i}=1,\ \mbox{and}\ \sum_{i=1}^{n}q_{i}\left(\begin{array}{l}
R_{i}S_{i}\boldsymbol{s}_{cc,\beta}(Y_{i},\boldsymbol{X}_{i},\boldsymbol{Z}_{i};\boldsymbol{\beta},\boldsymbol{\alpha})\\
R_{i}S_{i}\boldsymbol{v}(\boldsymbol{X}_{i},\boldsymbol{Z}_{i};\boldsymbol{\beta},\boldsymbol{\alpha},\boldsymbol{\theta})\\
R_{i}S_{i}\boldsymbol{s}_{cc,\alpha}(Y_{i},\boldsymbol{X}_{i},\boldsymbol{Z}_{i};\boldsymbol{\beta},\boldsymbol{\alpha})\\
S_{i}\boldsymbol{s}_{\alpha}(Y_{i},\boldsymbol{X}_{i},R_{i};\boldsymbol{\alpha})\\
\boldsymbol{h}(Y_{i},\boldsymbol{X}_{i};\boldsymbol{\theta})
\end{array}\right)={\bf 0},
\end{align*}
and $\widehat{\boldsymbol{\beta}}_{\text{\ensuremath{\widehat{\pi}\widehat{\theta}}-5}}$
can be redefined by replacing the third set of constraints above by
\begin{align*}
\sum_{i=1}^{n}q_{i}\left(\begin{array}{l}
R_{i}S_{i}\boldsymbol{s}_{cc,\beta}(Y_{i},\boldsymbol{X}_{i},\boldsymbol{Z}_{i};\boldsymbol{\beta},\boldsymbol{\alpha})\\
R_{i}S_{i}\boldsymbol{v}(\boldsymbol{X}_{i},\boldsymbol{Z}_{i};\boldsymbol{\beta},\boldsymbol{\alpha},\boldsymbol{\theta})\\
S_{i}\boldsymbol{s}_{\alpha}(Y_{i},\boldsymbol{X}_{i},R_{i};\boldsymbol{\alpha})-R_{i}S_{i}\boldsymbol{s}_{cc,\alpha}(Y_{i},\boldsymbol{X}_{i},\boldsymbol{Z}_{i};\boldsymbol{\beta},\boldsymbol{\alpha})\\
\boldsymbol{h}(Y_{i},\boldsymbol{X}_{i};\boldsymbol{\theta})
\end{array}\right)={\bf 0}.
\end{align*}
In the absence of multicollinearity among the components of $\boldsymbol{s}_{cc,\alpha}(Y_{i},\boldsymbol{X}_{i},\boldsymbol{Z}_{i};\boldsymbol{\beta},\boldsymbol{\alpha})$,
$\widehat{\boldsymbol{\beta}}_{\text{\ensuremath{\widehat{\pi}\widehat{\theta}}-4}}$
achieves the maximum efficiency based on the estimating functions
under consideration. But $\widehat{\boldsymbol{\beta}}_{\text{\ensuremath{\widehat{\pi}\widehat{\theta}}-5}}$
avoids the potential multicollinearity and still has a high efficiency.
We again recommend $\widehat{\boldsymbol{\beta}}_{\text{\ensuremath{\widehat{\pi}\widehat{\theta}}-5}}$
for use. The asymptotic variance for $\widehat{\boldsymbol{\beta}}_{\text{\ensuremath{\widehat{\pi}\widehat{\theta}}-5}}$
can again be derived by directly applying the result in \cite{qin1994empirical}. 

The constrained maximizations for all different estimators can be solved by the Lagrange
multiplier method, which has been well documented in the literature,
including \cite{qin1994empirical} and \cite{owen2001empirical}. Based on prior literature,
\cite{han2019empirical} provides a detailed algorithm that can be directly
applied to implement the estimators proposed in this article. Caveats
and potential numerical issues discussed there also apply. Readers
are referred to \cite{han2019empirical} for details about numerical
implementation algorithms. 

\section{Simulation Studies}

\subsection{Logistic regression with an expensive covariate}

In this study the binary outcome $Y$ is generated from $P(Y=1|X,Z;\boldsymbol{\beta})=\text{expit}(\beta_{c}+\beta_{X}X+\beta_{Z}Z)$.
For the covariates, we first generate $(\widetilde{X},Z)$ from a
bivariate normal distribution with standard normal marginals and correlation
coefficient $0.1$, and then let $X$ be a categorization of $\widetilde{X}$
with value 0, 1 or 2, corresponding to values of $\widetilde{X}$
falling into the three intervals $(-\infty,-0.44]$, $(-0.44,0.44]$
and $(0.44,\infty)$, respectively, determined by the tertiles $-0.44$
and $0.44$ for $\widetilde{X}$. The Phase 2 sampling probability
is $\pi(Y,\boldsymbol{X};\boldsymbol{\alpha})=\text{expit}(\alpha_{c}+\alpha_{Y}Y)$.
In our simulation we set $\boldsymbol{\beta}_{0}=(\beta_{c},\beta_{X},\beta_{Z})=(-4,1,1)$,
and $\boldsymbol{\alpha}_{0}=(\alpha_{c},\alpha_{Y})=(-3.5,2.3)$,
which leads to about $10\%$ of Phase 1 subjects with $Y=1$, about
$5\%$ of Phase 1 subjects entering into Phase 2, and Phase 2 sample
having roughly equal number of $Y=1$ and $Y=0$ outcomes. 

For the working model for $P(Y=1|X)$ we fit a logistic regression
$\text{expit}(\theta_{c}+\theta_{X}X)$ and take $\boldsymbol{h}(Y,X;\boldsymbol{\theta})$
to be the corresponding score function. For the selection probability
$\pi(Y,X)$ we fit the model $\pi(Y,X;\boldsymbol{\alpha})$. We also
fit a post-stratification model for $\pi(Y,X)$, $\text{logit}(\pi(Y,X;\boldsymbol{\alpha}))=\alpha_{c}+\alpha_{Y}Y+\alpha_{X1}I(X=1)+\alpha_{X2}I(X=2)$.
Post-stratification is commonly used in analyzing two-phase design
study data to improve the estimation efficiency (e.g.\cite{lawless1999semiparametric,scott2011fitting}). It works the same way as augmenting the
correct missingness model in the missing data literature (e.g. \cite{robins1994estimation}). 

\begin{scriptsize}
\ctable[
caption = {Simulation results with a binary outcome with an expensive covariate based on $1000$ replications. All Phase 1 subjects have positive selection probabilities. The numbers are biases, with empirical standard errors in $(\;)$, average estimated standard errors in $\{\;\}$, and coverage rate in $\langle\; \rangle$.}
]
{lllll}{
	\tnote[]{-$\pi$: using the true selection probabilities. -$\hat{\pi}$: using the estimated selection probabilities. -ps: using the estimated selection probabilities based on post-stratification.}
}{
	\multicolumn{4}{c}{Phase 1 sample size $n=2000$ (Phase 2 sample size about $100$)}\\
	&$\beta_{c}=-4$  & $\beta_{Z}=1$  & $\beta_{X}=1$  \\
	MLE	& -0.0700 (0.3688)&0.0485 (0.3158)&0.0278 (0.1977)\\
  	PS	& -0.0303 (0.3640)&0.0122 (0.3084)&0.0179 (0.2025)\\
	PS-ps	& -0.0298 (0.3578)&-0.0105 (0.3060)&0.0179 (0.1988)\\
	CML-$\pi$ &	-0.1265 (0.5971)&0.0515 (0.3131)&0.0603 (0.3574)\\
	CML-$\hat{\pi}$ &	-0.1187 (0.5473)&0.0515 (0.3131)&0.0603 (0.3574)\\
	CML-ps	& -0.1173 (0.5442)&0.0514 (0.3131)&0.0581 (0.3557)\\
	SW	& -0.1173 (0.5443)&0.0514 (0.3131)&0.0580 (0.3557)\\
	SW-ps	& -0.1187 (0.5473)&0.0515 (0.3131)&0.0603 (0.3574)\\
	EL-$\hat{\pi}\hat{\theta}$-5	& -0.0763 (0.3829)&0.0562 (0.3181)&0.0305 (0.2048)\\
	&$\langle0.9380\rangle$ \{0.3476\}&$\langle 0.9470\rangle$ \{0.2965\}&$\langle 0.9580\rangle$ \{0.1953\}\\
	EL-$\hat{\pi}\hat{\theta}$-5-ps	& -0.1600 (0.4031)&0.0323 (0.3242)&0.0621 (0.2153)\\
	&$\langle0.9450\rangle$ \{0.3611\}&$\langle0.9330\rangle$ \{0.2952\}&$\langle0.9500\rangle$ \{0.2000\}\\
	\vspace{2mm} & & & & \\
	\multicolumn{4}{c}{Phase 1 sample size $n=8000$ (Phase 2 sample size about $400$)}\\
	&$\beta_{c}=-4$  & $\beta_{Z}=1$  & $\beta_{X}=1$  \\
	MLE                   &     -0.0162 (0.1661)&0.0066 (0.1416)&0.0078 (0.0913)\\
    PS	& -0.0053 (0.1756)&-0.0108 (0.1429)&0.0048 (0.1005)\\
	PS-ps	& -0.0058 (0.1743)&-0.0072 (0.1406)&0.0047 (0.1006)\\
    CML-$\pi$      &	-0.0305 (0.2654)&0.0082 (0.1418)&0.0177 (0.1648)\\
	CML-$\hat{\pi}$ &     -0.0313 (0.2466)&0.0082 (0.1418)&0.0177 (0.1648)\\
	CML-ps	         &      -0.0303 (0.2454)&0.0082 (0.1418)&0.0175 (0.1640)\\
	SW	                  &-0.0313 (0.2466)&0.0082 (0.1418)&0.0177 (0.1648)\\
	SW-ps	            & -0.0302 (0.2455)&0.0082 (0.1418)&0.0176 (0.1640)\\
    EL-$\hat{\pi}\hat{\theta}$-5                &-0.0157 (0.1669)&0.0082 (0.1419)&0.0076 (0.0916)\\
	&$\langle0.9420\rangle$ \{0.1658\} &$\langle0.9450\rangle$ \{0.1415\}&$\langle0.9500\rangle$ \{0.0929\}\\
	EL-$\hat{\pi}\hat{\theta}$-5-ps           	&-0.0940 (0.1931)&0.0038 (0.1468)&0.0394 (0.1019)\\
	&$\langle0.9120\rangle$ $\{0.1731\}$&$\langle0.9360\rangle$ \{0.1425\}&$\langle0.9200\rangle$ \{0.0956\}\\
}
\end{scriptsize}

Table 1 contains the simulation results based on $1000$ replications.
Because $X$ is categorical, we are able to compute the maximum likelihood
estimator (MLE) of $\boldsymbol{\beta}_{0}$ following \cite{zhao2009likelihood} and the pseudoscore estimator (PS) following \cite{chatterjee2003pseudoscore} by modeling $f(Z\mid X)$ nonparametrically, and these two methods are
included as the benchmark for comparison. The estimator (SW) in \cite{scott2011fitting} defined by solving (\ref{SW}) is also included. Our
proposed estimator is termed as EL because of its empirical-likelihood-based
implementation. Under this simulation setting, it can be shown that
some components of $\boldsymbol{s}_{c,\alpha}(Y,\boldsymbol{X},\boldsymbol{Z};\boldsymbol{\beta},\boldsymbol{\alpha})$
in (\ref{EL4}) are linearly dependent, and thus we only compute $\widehat{\boldsymbol{\beta}}_{\text{\ensuremath{\widehat{\pi}\widehat{\theta}}-5}}$. 

From Table 1, compared to both the CML and the SW estimators, $\widehat{\boldsymbol{\beta}}_{\text{\ensuremath{\widehat{\pi}\widehat{\theta}}-5}}$
has substantially smaller empirical standard errors for $\beta_{c}$
and $\beta_{X}$, corresponding to the covariates used in the working
model for $P(Y=1|X)$. The empirical standard errors of $\widehat{\boldsymbol{\beta}}_{\text{\ensuremath{\widehat{\pi}\widehat{\theta}}-5}}$
without post-stratification are very close to that of the MLE, showing
the high efficiency of the proposed estimator. After post-stratification,
the empirical standard errors of $\widehat{\boldsymbol{\beta}}_{\text{\ensuremath{\widehat{\pi}\widehat{\theta}}-5}}$
become larger, and this is largely because the post-stratification
model for $\pi(Y,X)$ increases the number of constraints in (\ref{EL5})
and thus may jeopardize the numerical performance. With Phase 1 sample
size $n=2000$ (Phase 2 sample size about $100$), the average estimated
standard errors for $\widehat{\boldsymbol{\beta}}_{\text{\ensuremath{\widehat{\pi}\widehat{\theta}}-5}}$
are smaller than the empirical standard errors, but the difference
becomes negligible as the Phase 1 sample size increases to $n=8000$
(Phase 2 sample size about $400$).  

\subsection{Logistic regression with a surrogate covariate}

In this study the binary outcome $Y$ is generated from $P(Y=1|X,Z;\boldsymbol{\beta})=\text{expit}(\beta_{c}+\beta_{Z}Z)$, so only $Z$ influences the outcome. $X$ is assumed to be a surrogate of $Z$. To achieve this, we generate $({X},Z)$ from a
bivariate normal distribution with standard normal marginals and correlation
coefficient $\rho$, which is varied from 0.9 to 0.7 to simulation different degrees of similarity between the surrogate and real influential covariate. The Phase 2 sampling probability
is $\pi(Y,\boldsymbol{X};\boldsymbol{\alpha})=\text{expit}(\alpha_{c}+\alpha_{Y}Y)$.
In our simulation we set $\boldsymbol{\beta}_{0}=(\beta_{c},\beta_{Z})=(-3.3,1)$,
and $\boldsymbol{\alpha}_{0}=(\alpha_{c},\alpha_{Y})=(-3.5,3.5)$,
which leads to about $5\%$ of Phase 1 subjects with $Y=1$, about
$5\%$ of Phase 1 subjects entering into Phase 2, and Phase 2 sample
having roughly equal number of $Y=1$ and $Y=0$ outcomes. 

For the working model for $P(Y=1|X)$ we fit a logistic regression
$\text{expit}(\theta_{c}+\theta_{X}X)$ and take $\boldsymbol{h}(Y,X;\boldsymbol{\theta})$
to be the corresponding score function. For the selection probability
$\pi(Y,X)$ we fit the model $\pi(Y,X;\boldsymbol{\alpha})$. We also
fit a post-stratification model for $\pi(Y,X)$, $\text{logit}(\pi(Y,X;\boldsymbol{\alpha}))=\alpha_{c}+\alpha_{Y}Y+\alpha_{X}X$.

\begin{scriptsize}
\ctable[
caption = {Simulation results with a binary outcome and a surrogate covariate based on $1000$ replications. All Phase 1 subjects have positive selection probabilities. The numbers are biases, with empirical standard errors in $(\;)$, average estimated standard errors in $\{\;\}$, and coverage rate in $\langle\; \rangle$.}
]
{lcccc}{
\label{log_sur}
	\tnote[]{-$\pi$: using the true selection probabilities. -$\hat{\pi}$: using the estimated selection probabilities. -ps: using the estimated selection probabilities based on post-stratification.}
}{
	\multicolumn{5}{c}{Phase 1 sample size $n=2000$ (Phase 2 sample size about 100)}\\
	&\multicolumn{2}{c}{$\rho=0.9$}&\multicolumn{2}{c}{$\rho=0.7$}\\
	&$\beta_{c}=-3.3$   &  $\beta_Z=1$&$\beta_{c}=-3.3$   &  $\beta_Z=1$ \\
	%MLE	& -0.0700 (0.3688)&0.0485 (0.3158)\\
	CML-$\pi$ &	-0.0107 (0.2382)&0.0315 (0.2557)&-0.0107 (0.2382)&0.0315 (0.2557)\\
	CML-$\hat{\pi}$ &	-0.0130 (0.1763)&0.0315 (0.2557)&	-0.0130 (0.1763)&0.0315 (0.2557)\\
	CML-ps	& -0.0090 (0.1580)&0.0189 (0.2158)&-0.0115 (0.1671)& 0.0251 (0.2404)\\
	SW	& -0.00130 (0.1763)&0.0315 (0.2557)&-0.0130 (0.1763)&-0.0315 (0.2557) \\
	SW-ps	& -0.0248 (0.2228)&0.0408 (0.3295)&-0.0202 (0.1996)&0.0376 (0.2908)\\
   
	EL-$\hat{\pi}\hat{\theta}$-5	& -0.0063 (0.1410)&0.0025 (0.1422)& -0.0102 (0.1570)&0.0141 (0.1933) \\
	&$\langle0.9580\rangle$ \{0.1442\}&$\langle 0.9310\rangle$ \{0.1332\}&$\langle0.9490\rangle$ \{0.1586\}&$\langle 0.9300\rangle$ \{0.1772\}\\
	EL-$\hat{\pi}\hat{\theta}$-5-ps	& -0.0199 (0.1435)& 0.0007 (0.1471)	& -0.0255 (0.1596)& 0.0158 (0.1984)\\
    &$\langle0.9570\rangle$ \{0.1464\}&$\langle 0.9230\rangle$ \{0.1349\} &$\langle0.9480\rangle$ \{0.1611\}&$\langle 0.9270\rangle$ \{0.1793\}\\
	\vspace{2mm} & &  \\
	\multicolumn{5}{c}{Phase 1 sample size $n=8000$ (Phase 2 sample size about 400)}\\
	&\multicolumn{2}{c}{$\rho=0.9$}&\multicolumn{2}{c}{$\rho=0.7$}\\
	&$\beta_{c}=-3.3$   &  $\beta_Z=1$&$\beta_{c}=-3.3$   &  $\beta_Z=1$ \\
	%MLE	& -0.0700 (0.3688)&0.0485 (0.3158)\\
	CML-$\pi$ &	0.0001 (0.1205)&0.0010 (0.1217) &	0.0001 (0.1205)&0.0010 (0.1217)\\
	CML-$\hat{\pi}$ &	-0.0040 (0.0859)&0.0010 (0.1217)&	-0.0040 (0.0859)&0.0010 (0.1217)\\
	CML-ps	& -0.0044 (0.0773)&-0.0002 (0.0983)& -0.0042 (0.0811)&-0.0003 (0.1100)\\
	SW	& -0.00040 (0.0859)&0.0010 (0.1217)& -0.00040 (0.0859)&0.0010 (0.1217)\\
	SW-ps	& -0.0042 (0.0827)&-0.0003 (0.1198)	& -0.0048 (0.0874)&0.0005 (0.1243)\\
   
	EL-$\hat{\pi}\hat{\theta}$-5	& -0.0022 (0.0716)&-0.0041 (0.0668)	& -0.0027 (0.0777)&-0.0045 (0.0884)\\
	&$\langle0.9550\rangle$ \{0.0723\}&$\langle 0.9470\rangle$ \{0.0647\}	&$\langle0.9560\rangle$ \{0.0788\}&$\langle 0.9500\rangle$ \{0.0884\}\\
	EL-$\hat{\pi}\hat{\theta}$-5-ps	& -0.0063 (0.0720)& -0.0049 (0.0675)	& -0.0081 (0.0783)& -0.0046 (0.0902)\\
    &$\langle0.9540\rangle$ \{0.0726\}&$\langle 0.9480\rangle$ \{0.0676\}&$\langle0.9540\rangle$ \{0.0792\}&$\langle 0.9500\rangle$ \{0.0887\}\\
	\vspace{2mm} & &  \\
%	\multicolumn{3}{c}{Phase 1 sample size $n=2000$, $\rho=0.7$}\\
%	&$\beta_{c}=-3.3$   &  $\beta_Z=1$\\
%	%MLE	& -0.0700 (0.3688)&0.0485 (0.3158)\\
%	CML-$\pi$ & 0.0251 (0.2920)&0.0343 (0.2586)\\
%	CML-$\hat{\pi}$ &	0.0305 (0.2420)&0.0343 (0.2586)\\
%	CML-ps	& 0.0334 (0.2365)&0.0299 (0.2380)\\
%	SW	& 0.0305 (0.2420)&0.0343 (0.2586)\\
%	SW-ps	& 0.0250 (0.2571)&0.0391 (0.2813)\\
   
%	EL-$\hat{\pi}\hat{\theta}$-5	& 0.0351 (0.2316)&0.0181 (0.1911)\\
%	&$\langle0.8510\rangle$ \{0.1580\}&$\langle 0.9360\rangle$ \{0.1778\}\\
%	EL-$\hat{\pi}\hat{\theta}$-5-ps	& 0.0212 (0.2320)& 0.0183(0.1931)\\
%   &$\langle0.8540\rangle$ \{0.1605\}&$\langle 0.9330\rangle$ \{0.1797\}\\
}
\end{scriptsize}

Table \ref{log_sur} contains the simulation results based on $1000$ replications.
Because $X$ is continuous, the maximum likelihood
estimator (MLE) of $\boldsymbol{\beta}_{0}$ following \cite{zhao2009likelihood} and the pseudoscore estimator (PS) following \cite{chatterjee2003pseudoscore} are no longer feasible to compute. Compared to both the CML and the SW estimators, we see that for a surrogate covariate which has high correlation with the influential covariate $Z$, the EL estimators significantly improved the estimation efficiency. The improvement decreases as the correlation between the surrogate and original covariate decreases, which is sensible.

\subsection{Linear regression with an expensive covariate}

In this study the outcome $Y$ follows a normal distribution $f(Y|X,Z;\boldsymbol{\beta})=N(\beta_{c}+\beta_{X}X+\beta_{Z}Z,\beta_{\sigma^{2}})$,
where $\beta_{\sigma^{2}}$ is the variance. The covariates $X$ and
$Z$ are generated in the same way as in Section 4.1. For Phase 2
selection, we consider a sampling plan often used in genetic epidemiology,
in which only the lower and upper tails of the $Y$ distribution are sampled.
The sampling probability is $\pi(Y,X;\boldsymbol{\alpha})=\alpha_{1}I(Y\in S_{1})+\alpha_{2}I(Y\in S_{2})$,
where $S_{1}$ and $S_{2}$ are the two strata $(-\infty,-0.63]$ and $(2.63,\infty)$ determined by the first and third
quartiles, $-0.63$ and $2.63$, for $Y$. Thus subjects with
$Y$ values in $(-0.63,2.63]$ have a zero selection probability. We take $\boldsymbol{\beta}_{0}=(\beta_{c},\beta_{X},\beta_{Z},\beta_{\sigma^{2}})=(0,1,1,2^{2})$ and $\boldsymbol{\alpha}_{0}=(0.3,0.5)$, leading to about $20\%$ of the Phase 1 subjects being selected to enter Phase 2. For the working model
for $f(Y|X)$ we fit a linear regression that leads to $\boldsymbol{h}(Y,X;\boldsymbol{\theta})=((Y-\theta_{c}-\theta_{X}X),(Y-\theta_{c}-\theta_{X}X)X)^{T}$. For $\pi(Y,X)$ we fit the model $\pi(Y,X;\boldsymbol{\alpha})$
with $\boldsymbol{\alpha}=(\alpha_{1},\alpha_{2})$. We also fit a
post-stratification model for $\pi(Y,X)$ by including $X$. The post-stratification
model has a six dimensional $\boldsymbol{\alpha}$ that allows a different
selection probability for each of the six strata defined by the three
categories of $X$ and the strata $S_{1}$ and $S_{2}$ for $Y$.

\begin{scriptsize}
\ctable[
caption = {Simulation results with a continuous outcome with an expensive covariate based on $1000$ replications. Some Phase 1 subjects have a zero selection probability. The numbers are biases, with empirical standard errors in $(\;)$, average estimated standard errors in $\{\;\}$, and coverage rate in $\langle\; \rangle$. } 
]
{lllll}{
	\tnote[]{-$\pi$: using the true selection probabilities. -$\hat{\pi}$: using the estimated selection probabilities. -ps: using the estimated selection probabilities based on post-stratification.}
}{
	\multicolumn{5}{c}{Phase 1 sample size $n=300$ (Phase 2 sample size about $60$)}\\
	& $\beta_{c}=0$  & $\beta_{Z}=1$  & $\beta_{X}=1$ &$\sigma=2$ \\
	MLE	& \ 0.0068 (0.2626)&-0.0021 (0.1965)& 0.0357 (0.2223)&-0.0239 (0.1154)\\
     PS & -0.0044 (0.2272)& -0.0173 (0.1896) & 0.0022 (0.1785) & 0.0025 (0.0943)\\
    PS-ps& -0.0012 (0.2271)& -0.0025 (0.1881) & 0.0023 (0.0764) & -0.0043 (0.0948)\\
	CML-$\pi$ &-0.0154 (0.3606)& 0.0319 (0.2355)& 0.0213 (0.2723)&-0.0369 (0.1611)\\
	CML-$\hat{\pi}$&-0.0141 (0.3391)& 0.0350 (0.2359)& 0.0249 (0.2736)&-0.0340 (0.1617)\\
	CML-ps& -0.0005 (0.3052)& 0.0418 (0.2389)& 0.0061 (0.2363)&-0.0283 (0.1611)\\
	SW& -0.0142 (0.3381)& 0.0353 (0.2361)& 0.0252 (0.2737)&-0.0338 (0.1617)\\
	SW-ps& -0.0039 (0.3061)& 0.0402 (0.2402)& 0.0091 (0.2306)&-0.0276 (0.1604)\\
	EL-$\hat{\pi}\hat{\theta}$-4	&\ 0.0076 (0.2620)&\ 0.0377 (0.2437)&-0.0026 (0.1985)&-0.0347 (0.1691)\\
	&$\langle 0.9290\rangle$\{0.2564\}&$\langle 0.9060\rangle$\{0.2216\} &$\langle0.9440\rangle$\{0.1955\} &$\langle0.8860\rangle$\{0.1412\} \\
	EL-$\hat{\pi}\hat{\theta}$-4-ps	&\ 0.0028 (0.2798)&\ 0.0421 (0.2449)&\ 0.0011 (0.2076)&-0.0433 (0.1610)\\
	&$\langle0.9270\rangle$\{0.2504\} &$\langle0.9250\rangle$\{0.2120\} &$\langle0.9530\rangle$\{0.1910\} &$\langle0.9190\rangle$\{0.1320\} \\
	EL-$\hat{\pi}\hat{\theta}$-5	&\ 0.0068 (0.2646)&\ 0.0421 (0.2421)&-0.0025 (0.2000)&-0.0275 (0.1662)\\
	&$\langle0.9380\rangle$\{0.2596\} &$\langle0.9250\rangle$\{0.2257\} &$\langle0.9530\rangle$\{0.1977\} &$\langle0.9190\rangle$\{0.1459\} \\
	EL-$\hat{\pi}\hat{\theta}$-5-ps	&\ 0.0056 (0.2774)&\ 0.0416 (0.2365)&-0.0012 (0.2044)&-0.0257 (0.1542)\\
	\vspace{2mm} &$\langle0.9390\rangle$\{0.2592\} &$\langle0.9620\rangle$\{0.2244\} &$\langle0.9040\rangle$\{0.1966\} &$\langle0.9640\rangle$\{0.1417\} \\
	\multicolumn{5}{c}{Phase 1 sample size $n=2000$ (Phase 2 sample size about $400$)}\\
	& $\beta_{c}=0$  & $\beta_{Z}=1$  & $\beta_{X}=1$ &$\sigma=2$ \\
	MLE	&\ 0.0005 (0.1014)&-0.0023 (0.0757)& 0.0029 (0.0802)&-0.0031 (0.0434)\\
	PS & 0.0012 (0.1053)& -0.0071 (0.0797) & -0.0046 (0.0794) & 0.0018 (0.0431)\\
    PS-ps& 0.0006 (0.1028)& -0.0038 (0.0797) & -0.0031 (0.0764) & 0.0001 (0.0430)\\
    CML-$\pi$ &	-0.0014 (0.1336)&-0.0006 (0.0897)& 0.0024 (0.1068)&-0.0070 (0.0587)\\
	CML-$\hat{\pi}$&-0.0016 (0.1267)&-0.0001 (0.0898)& 0.0029 (0.1069)&-0.0065 (0.0587)\\
	CML-ps&\  0.0008 (0.1137)& 0.0011 (0.0891)&-0.0015 (0.0894)&-0.0053 (0.0579)\\
	SW& -0.0014 (0.1257)&-0.0001 (0.0898)& 0.0029 (0.1069)&-0.0065 (0.0587)\\
	SW-ps&\ 0.0010 (0.1111)& 0.0012 (0.0890)&-0.0018 (0.0869)&-0.0051 (0.0579)\\
     
	EL-$\hat{\pi}\hat{\theta}$-4	&\  0.0010 (0.1019)&\ 0.0014 (0.0905)&-0.0023 (0.0769)&-0.0050 (0.0584)\\
	&$\langle0.9530\rangle$\{0.0994\}&$\langle0.9530\rangle$\{0.0896\}&$\langle0.9460\rangle$\{0.0755\}&$\langle 0.9610 \rangle$\{0.0591\}\\
	EL-$\hat{\pi}\hat{\theta}$-4-ps	&\ 0.0011 (0.1025)&\ 0.0016 (0.0877)&-0.0025 (0.0769)&-0.0049 (0.0550)\\
	&$\langle 0.9540 \rangle$\{0.0991\}&$\langle0.9370 \rangle$\{0.0872\}&$\langle 0.9470 \rangle$\{0.0753\}&$\langle0.9310 \rangle$\{0.0559\}\\
	EL-$\hat{\pi}\hat{\theta}$-5	&\ 0.0013 (0.1019)&\ 0.0016 (0.0901)&-0.0025 (0.0768)&-0.0051 (0.0583)\\
	&$\langle0.9580 \rangle$\{0.0996\}&$\langle0.9570 \rangle$\{0.0898\}&$\langle0.9480 \rangle$\{0.0756\}&$\langle0.9580 \rangle$\{0.0593\}\\
	EL-$\hat{\pi}\hat{\theta}$-5-ps	&\  0.0013 (0.1019)&\ 0.0018 (0.0864)&-0.0027 (0.0766)&-
	0.0045 (0.0550)\\
	&$\langle0.9590 \rangle$\{0.0995\}&$\langle0.9800 \rangle$\{0.0880\}&$\langle0.9220 \rangle$\{0.0756\}&$\langle0.9830 \rangle$\{0.0565\}\\
}
\end{scriptsize}

Table 3 contains the simulation results based on $1000$ replications. It is seen by comparing the empirical standard errors that, without
post-stratification, the proposed estimators $\widehat{\boldsymbol{\beta}}_{\text{\ensuremath{\widehat{\pi}\widehat{\theta}}-4}}$
and $\widehat{\boldsymbol{\beta}}_{\text{\ensuremath{\widehat{\pi}\widehat{\theta}}-5}}$
have considerable efficiency improvements over both the CML and the
SW estimators in $\beta_{c}$ and $\beta_{X}$, the covariates used
in the model for $f(Y|X)$. With post-stratification, the SW estimator
has a very good efficiency but still has larger empirical standard
errors on average compared to $\widehat{\boldsymbol{\beta}}_{\text{\ensuremath{\widehat{\pi}\widehat{\theta}}-4}}$
and $\widehat{\boldsymbol{\beta}}_{\text{\ensuremath{\widehat{\pi}\widehat{\theta}}-5}}$.
The $\widehat{\boldsymbol{\beta}}_{\text{\ensuremath{\widehat{\pi}\widehat{\theta}}-4}}$
and $\widehat{\boldsymbol{\beta}}_{\text{\ensuremath{\widehat{\pi}\widehat{\theta}}-5}}$
have very similar numerical performance, and post-stratification does
not help much for further efficiency improvement and actually increases
the empirical standard errors when the sample size is not large ($n=300$). Inspecting the average estimated standard errors, we see that for
$\widehat{\boldsymbol{\beta}}_{\text{\ensuremath{\widehat{\pi}\widehat{\theta}}-4}}$
the standard errors are underestimated, especially with post-stratification
and a small sample size ($n=300$). This may be largely due to the
increased number of constraints, which can jeopardize the numerical
performance. For $\widehat{\boldsymbol{\beta}}_{\text{\ensuremath{\widehat{\pi}\widehat{\theta}}-5}}$
the estimated standard errors are close to the empirical standard
errors.  
\subsection{Linear regression with a surrogate covariate}

In this study the outcome $Y$ follows a normal distribution $f(Y|X,Z;\boldsymbol{\beta})=N(\beta_{c}+\beta_{Z}Z,\beta_{\sigma^{2}})$,
where $\beta_{\sigma^{2}}$ is the variance. The covariates $X$ and
$Z$ are generated in the same way as in Section 4.2, with correlation coefficient varying from 0.7 to 0.9. For Phase 2
selection, the sampling plan is similar to Section 4.3 in which only the lower and upper tails of the $Y$ distribution are sampled. We take $\boldsymbol{\beta}_{0}=(\beta_{c},\beta_{Z},\beta_{\sigma^{2}})=(0,1,2^{2})$. The two strata $S_{1}$ and $S_{2}$ are similarly determined by the first and third
quartiles for $Y$ but with different cut points as the distribution of $Y$ is different now. In this scenario we use $S_{1}=(-\infty,-1.52]$ and $S_2=(1.52,\infty)$. 
$Y$ values in $(-1.52,1.52]$ have a zero selection probability. $\boldsymbol{\alpha}_{0}=(0.3,0.5)$ leading to about $20\%$ of the Phase 1 subjects being selected to enter Phase 2. For the working model for $f(Y|X)$ we fit a linear regression that leads to $\boldsymbol{h}(Y,X;\boldsymbol{\theta})=((Y-\theta_{c}-\theta_{X}X),(Y-\theta_{c}-\theta_{X}X)X)^{T}$. The post-stratification model for $\pi(Y,X)$ by stratifying on 3 tertiles of $X$, which leads to a six dimensional $\boldsymbol{\alpha}$.

Results of 1000 replications are shown in Table 4 and 5. We observe similar numerical performances of the 4 EL estimators. Similar to the results in Section 4.3, when the sample size is small ($n=300$, with about 20\% into Phase 2), the post-stratification no longer leads to further efficiency improvement, likely due to the jeopardized numerical performance. The efficiency improvement on $\beta_Z$ decreases as the correlation between the surrogate and the true influential covariate decreases. Yet with moderate correlation $\rho=0.7$, we still observe substantial efficiency improvement, especially when $n$ is small. This is of significance, as in the surrogate covariate scenario, the only efficiency improvement of CML and SW relies on post-stratification, which can be a problem when $n$ is small.

\begin{scriptsize}
\ctable[
caption = {Simulation results with a continuous outcome and a surrogate covariate based on $1000$ replications. Some Phase 1 subjects have a zero selection probability. The numbers are biases, with empirical standard errors in $(\;)$, average estimated standard errors in $\{\;\}$, and coverage rate in $\langle\; \rangle$. } 
]
{lccc}{
	\tnote[]{-$\pi$: using the true selection probabilities. -$\hat{\pi}$: using the estimated selection probabilities. -ps: using the estimated selection probabilities based on post-stratification.}
}{
	\multicolumn{4}{c}{Phase 1 sample size $n=300$, $\rho=0.9$}\\
	& $\beta_{c}=0$  & $\beta_{Z}=1$  &$\sigma=2$ \\
	CML-$\pi$ &0.0050 (0.2315)& 0.0343 (0.2419)& -0.0204 (0.1630)\\
	CML-$\hat{\pi}$&0.0094 (0.1856)& 0.0383 (0.2422)& -0.0167 (0.1634)\\
	CML-ps& 0.0074 (0.1749)& 0.0315 (0.2207)& -0.0118 (0.1603)\\
	SW& 0.0089 (0.1813)& 0.0387 (0.2422)& -0.0163 (0.1633)\\
	SW-ps& 0.0059 (0.1629)& 0.0268 (0.2137)& -0.0084 (0.1583)\\
	EL-$\hat{\pi}\hat{\theta}$-4	&\ 0.0020 (0.1314)&\ 0.0037 (0.1545)&-0.0171 (0.1695)\\
	&$\langle 0.9420\rangle$\{0.1249\}&$\langle 0.9310\rangle$\{0.1384\} &$\langle0.8870\rangle$\{0.1410\}\\
	EL-$\hat{\pi}\hat{\theta}$-4-ps	&\ 0.0023 (0.1331)&\ 0.0061 (0.1585)&\ -0.0190 (0.1616)\\
	&$\langle0.9380\rangle$\{0.1233\} &$\langle0.9250\rangle$\{0.1367\} &$\langle0.8890\rangle$\{0.1336\} \\
	EL-$\hat{\pi}\hat{\theta}$-5	&\ 0.0016 (0.1313)&\ 0.0061 (0.1526)&-0.0080 (0.1648)\\
	&$\langle0.9460\rangle$\{0.1255\} &$\langle0.9370\rangle$\{0.1397\} &$\langle0.9110\rangle$\{0.1464\}\\
	EL-$\hat{\pi}\hat{\theta}$-5-ps	&\ 0.0015 (0.1359)&\ 0.0064 (0.1571)&-0.0043 (0.1578)\\
	\vspace{2mm} &$\langle0.9430\rangle$\{0.1253\} &$\langle0.9250\rangle$\{0.1407\} &$\langle0.9120\rangle$\{0.1414\} \\
	\multicolumn{4}{c}{Phase 1 sample size $n=300$, $\rho=0.7$}\\
	& $\beta_{c}=0$  & $\beta_{Z}=1$  &$\sigma=2$ \\
	CML-$\pi$ &0.0050 (0.2315)& 0.0343 (0.2419)& -0.0204 (0.1630)\\
	CML-$\hat{\pi}$&0.0094 (0.1856)& 0.0383 (0.2422)& -0.0167 (0.1634)\\
	CML-ps& 0.0079 (0.1808)& 0.0361 (0.2319)& -0.0144 (0.1631)\\
	SW& 0.0089 (0.1813)& 0.0387 (0.2422)& -0.0163 (0.1633)\\
	SW-ps& 0.0057 (0.1722)& 0.0305 (0.2273)& -0.0111 (0.1624)\\
	EL-$\hat{\pi}\hat{\theta}$-4	&\ 0.0021 (0.1523)&\ 0.0123 (0.2055)&-0.0195 (0.1699)\\
	&$\langle 0.9290\rangle$\{0.1402\}&$\langle 0.9280\rangle$\{0.1770\} &$\langle0.8910\rangle$\{0.1424\}\\
	EL-$\hat{\pi}\hat{\theta}$-4-ps	&\ 0.0009 (0.1524)&\ 0.0148 (0.2064)&\ -0.0209 (0.1676)\\
	&$\langle0.9210\rangle$\{0.1378\} &$\langle0.9280\rangle$\{0.1733\} &$\langle0.8900\rangle$\{0.1377\} \\
	EL-$\hat{\pi}\hat{\theta}$-5	&\ 0.0014 (0.1502)&\ 0.0107 (0.2036)&-0.0073 (0.1647)\\
	&$\langle0.9320\rangle$\{0.1412\} &$\langle0.9310\rangle$\{0.1795\} &$\langle0.9130\rangle$\{0.1476\}\\
	EL-$\hat{\pi}\hat{\theta}$-5-ps	&\ -0.0001 (0.1499)&\ 0.0107 (0.2046)&-0.0073 (0.1641)\\
	\vspace{2mm} &$\langle0.9340\rangle$\{0.1407\} &$\langle0.9230\rangle$\{0.1822\} &$\langle0.9150\rangle$\{0.1456\} \\
}
\ctable[
caption = {Simulation results with a continuous outcome and a surrogate covariate based on $1000$ replications. Some Phase 1 subjects have a zero selection probability. The numbers are biases, with empirical standard errors in $(\;)$, average estimated standard errors in $\{\;\}$, and coverage rate in $\langle\; \rangle$. } 
]
{lccc}{
	\tnote[]{-$\pi$: using the true selection probabilities. -$\hat{\pi}$: using the estimated selection probabilities. -ps: using the estimated selection probabilities based on post-stratification.}
}{
		\multicolumn{4}{c}{Phase 1 sample size $n=2000$, $\rho=0.9$}\\
	& $\beta_{c}=0$  & $\beta_{Z}=1$  &$\sigma=2$ \\
	CML-$\pi$ &-0.0009 (0.0800)& -0.0018 (0.0860)& -0.0041 (0.0598)\\
	CML-$\hat{\pi}$&-0.0011 (0.0658)& -0.0012 (0.0861)& -0.0035 (0.0598)\\
	CML-ps& -0.0019 (0.0612)& 0.0001 (0.0781)& -0.0034 (0.0585)\\
	SW& -0.0012 (0.0637)& -0.0011 (0.0861) & -0.0035 (0.0598)\\
	SW-ps& -0.0023 (0.0575)& -0.0003 (0.0738) & -0.0030 (0.0577)\\
	EL-$\hat{\pi}\hat{\theta}$-4	&\ -0.0025 (0.0483)&\ -0.0017 (0.0539)&-0.0025 (0.0593)\\
	&$\langle 0.9460\rangle$\{0.0489\}&$\langle 0.9580\rangle$\{0.0551\} &$\langle0.9500\rangle$\{0.0592\}\\
	EL-$\hat{\pi}\hat{\theta}$-4-ps	&\ -0.0024 (0.0483)&\ -0.0018 (0.0534)&\ -0.0031 (0.0559)\\
	&$\langle0.9490\rangle$\{0.0488\} &$\langle0.9590\rangle$\{0.0545\} &$\langle0.9530\rangle$\{0.0564\} \\
	EL-$\hat{\pi}\hat{\theta}$-5	&\ -0.0025 (0.0482)&\ -0.0016 (0.0537)&-0.0021 (0.0593)\\
	&$\langle0.9470\rangle$\{0.0489\} &$\langle0.9570\rangle$\{0.0551\} &$\langle0.9510\rangle$\{0.0595\}\\
	EL-$\hat{\pi}\hat{\theta}$-5-ps	&\ -0.0025 (0.0483)&\ -0.0018 (0.0537)&-0.0024 (0.0559)\\
	\vspace{2mm} &$\langle0.9470\rangle$\{0.0489\} &$\langle0.9600\rangle$\{0.0553\} &$\langle0.9510\rangle$\{0.0568\} \\
	\multicolumn{4}{c}{Phase 1 sample size $n=2000$, $\rho=0.7$}\\
	& $\beta_{c}=0$  & $\beta_{Z}=1$  &$\sigma=2$ \\
	CML-$\pi$ &-0.0009 (0.0800)& -0.0018 (0.0860)& -0.0041 (0.0598)\\
	CML-$\hat{\pi}$&-0.0011 (0.0658)& -0.0012 (0.0861)& -0.0035 (0.0598)\\
	CML-ps& -0.0018 (0.0634)& -0.0001 (0.0818)& -0.0035 (0.0592)\\
	SW& -0.0012 (0.0637)& -0.0011 (0.0861) & -0.0035 (0.0598)\\
	SW-ps& -0.0026 (0.0603)& -0.0015 (0.0783) & -0.0028 (0.0586)\\
	EL-$\hat{\pi}\hat{\theta}$-4	&\ -0.0030 (0.0541)&\ -0.023 (0.0684)&-0.0026 (0.0593)\\
	&$\langle 0.951\rangle$\{0.0546\}&$\langle 0.9640\rangle$\{0.0712\} &$\langle0.9450\rangle$\{0.0595\}\\
	EL-$\hat{\pi}\hat{\theta}$-4-ps	&\ -0.0028 (0.0540)&\ -0.0016 (0.0680)&\ -0.0030 (0.0571)\\
	&$\langle0.9520\rangle$\{0.0546\} &$\langle0.9620\rangle$\{0.0700\} &$\langle0.9500\rangle$\{0.0579\} \\
	EL-$\hat{\pi}\hat{\theta}$-5	&\ -0.0031 (0.0540)&\ -0.0021 (0.0681)&-0.0025 (0.0592)\\
	&$\langle0.9520\rangle$\{0.0546\} &$\langle0.9650\rangle$\{0.0713\} &$\langle0.9450\rangle$\{0.0597\}\\
	EL-$\hat{\pi}\hat{\theta}$-5-ps	&\ -0.0030 (0.0540)&\ -0.0026 (0.0683)&-0.0024 (0.0573)\\
	\vspace{2mm} &$\langle0.9520\rangle$\{0.0546\} &$\langle0.9690\rangle$\{0.0719\} &$\langle0.9470\rangle$\{0.0584\} \\
}
\end{scriptsize}

\section{An Illustration with the NHANES 2015-2016 study}
We consider an application of the proposed method to test 
associations between  body mass index (BMI, kg/m2) and systolic blood pressure (SBP, mmHg)  based on data from the  National Health and Nutrition Examination Survey (NHANES) 2015-2016. Our goal is to study the effect of BMI ($X_1$) on the SBP ($Y$). Blood pressure tends to increas with age \cite{baksi2009meta}, and is also thought to be associated with certain nutritional intakes, e.g., sodium \cite{hermansen2000diet} and BMI. BMI is also influenced by some nutritional intakes, e.g. fat. We add age ($X_2$) and dietary factors $\boldsymbol{Z}$ into our model to adjust for their effects on SBP and study the effect of BMI after this adjustment is done. Variables in $\boldsymbol{Z}$ include sodium (mg, Sodium), total saturated fat (g, SatFat) which are continuous, and Salt used in Preparation (SaltPrep) which is an ordinal variable, coded as 0 = Never, 1 = Rarely, 3 = Occasionally, 4 = Very Often. 
The dietary intake is measured through an 24-hour food recall interview by a trained interviewer, which estimates the types and amounts of foods and beverages consumed during the 24-hour period prior to the interview, thus estimates intakes of nutrients. The interviewers were required to take intense one-week training, and the interviews are reviewed by scientists to ensure reasonableness, consistency etc. Thus, the interview-based variables are harder to collect. NHANES took significant effort in data collection and a majority of participants have the dietary intake responses. This provides an opportunity to check the performance of our method, especially compared to maximum likelihood with fully observed data. In total, we have 6453 subjects with SBP, BMI, age and all the 3 dietary ingredient measurements. We mimic a two-phase study where the interview-based variables are obtained only for a phase 2 sub-sample. SBP is positive-valued and its distribution is skewed so we take the outcome $Y$ as log(SBP). By checking the plot of SBP vs BMI we find that taking the logarithm of BMI help to improve the linear relationship, so BMI is also log transformed. All variables including $Y$ and covariates $X$ and $Z$ are standardized to have mean zero. In addition, the 3 nutrients' intakes are also re-scaled to unit variance, as in their original units the coefficients' magnitude could be very small and the estimates could be affected by the numerical tolerances we set in the implementation.  The model of interest is then
\begin{equation*}
    \log (\mbox{SBP})=\beta_0+\beta_1\text{Sodium}+\beta_2\text{SatFat}+\beta_3\text{SaltPrep}+\beta_4\text{log(BMI)}+\beta_5\text{age},
\end{equation*}
and our primary interest lies in the estimation of $\beta_4$.

For the Phase 1 data on $Y$ and $\boldsymbol{X}=(\text{BMI},\text{age})$ we fit
a linear regression model and take $\boldsymbol{h}(Y,\boldsymbol{X};\boldsymbol{\theta})=((Y-\theta_{c}-\boldsymbol{\theta}_{X}^{T} \boldsymbol{X}),(Y-\theta_{c}-\boldsymbol{\theta}_{X}^T \boldsymbol{X})\boldsymbol{X}^{T})^{T}$. For phase 2 sampling, we stratify $Y$ into 3 strata $S_1=(-\infty,c_1], S_2=(c_1,c_2], S_3=(c_2,\infty)$ by its 1/4 and 3/4 quartile of the 6453 Phase 1 samples, with $c_1=-0.099$ and $c_2=0.086$, and sample only the first and last quartiles through Bernoulli sampling. The sampling probability we use for these two strata is $\boldsymbol{\alpha}_0=(0.4, 0.4)$. The Phase 2 sample then takes about one fifth of the Phase 1 sample. We show the results from a single randomly selected Phase 2 sample, for which 1279 individuals are selected through Bernoulli sampling.

For the selection probability $\pi(Y,\boldsymbol{X})$ we consider two models,
one is the correct model $\alpha_{1}I(Y\in S_{1})+\alpha_{3}I(Y\in S_{3})$,
and the other is a post-stratification model that allows speparate selection
probability for each of the eight strata determined by $S_{1}$
and $S_{3}$ for $Y$, and 4 quartiles $U_j, j=1,2,3,4$ of (standardized) BMI, with cutoff points -0.181, 0.003, 0.171.

Table 3 contains the analysis results with several different estimators for two-phase samples. As a comparison we also include
the results based on a full data analysis using MLE for all 6453 individuals. Our recommended estimator $\widehat{\boldsymbol{\beta}}_{\text{\ensuremath{\widehat{\pi}\widehat{\theta}}-5}}$
has a much improved efficiency compared to both the CML and the SW
estimates for the regression coefficients of log(BMI) and age.  With only 20\% of Phase 1 subjects selected to enter Phase 2, for the log(BMI) effect that is the primary interest, our recommended EL estimator yields estimate and p-value very close to those based on the MLE using the full data, and this shows the effectiveness of the proposed method. From Table 3, all methods indicate that BMI has a highly significant association with SBP, after adjusting for age and dietary variables. %The EL estimator achieves estimates and p-values very close to the model using the full data, whereas using only about 25\% of Phase 1 data. 

\begin{scriptsize}
\ctable[
caption ={Analysis results for the NHANES 2015-2016 data}
]
{llllllll}{
	\tnote[]{ MLE-F: MLE based on the whole Phase 1 data. -$\hat{\pi}$: using the estimated selection probabilities. -ps: using the estimated selection probabilities based on post-stratification.}
}{
	&& Intercept & Sodium & SatFat & SaltPrep & log(BMI) & age  \\

\multirow{3}{*}{MLE-F}& Estimate	& -8.87$\times 10^{-17}$ &5.04$\times 10^{-3}$  &-3.00$\times 10^{-3}$ &-4.33$\times 10^{-3}$ & 0.0959 & 0.0780\\
&S.E.&1.42$\times 10^{-3}$&2.00$\times 10^{-3}$&2.00$\times 10^{-3}$ &1.43$\times 10^{-3}$ & 5.89$\times 10^{-3}$ & 1.55$\times 10^{-3}$\\
&p-value&1&0.0118&0.134 &2.39$\times 10^{-3}$ & 0 & 0\\
\\
\multirow{3}{*}{CML-$\hat{\pi}$}& Estimate	& 4.75$\times 10^{-3}$ &5.41$\times 10^{-3}$  &-4.92$\times 10^{-3}$ & -4.87$\times 10^{-3}$ & 0.0953 & 0.0774\\
&S.E.&3.02$\times 10^{-3}$&4.57$\times 10^{-3}$&4.67$\times 10^{-3}$ & 3.15$\times 10^{-3}$ & 0.0123 & 3.81$\times 10^{-3}$\\
&p-value&0.11602&0.237&0.292 & 0.121 & 8.54$\times 10^{-15}$ & 0\\
\\
\multirow{3}{*}{CML-$\hat{\pi}$-ps}& Estimate	& 3.98$\times 10^{-3}$ &5.31$\times 10^{-3}$  &-4.77$\times 10^{-3}$ & -4.90$\times 10^{-3}$ & 0.0971 & 0.0778\\
&S.E.& 3.02$\times 10^{-3}$&4.57$\times 10^{-3}$&4.66$\times 10^{-3}$ & 3.14$\times 10^{-3}$ & 0.0124 & 3.83$\times 10^{-3}$\\
&p-value&0.188&0.245&0.306 & 0.118 & 3.93$\times 10^{-15}$ & 0\\
\\
\multirow{3}{*}{SW}& Estimate	& 2.17$\times 10^{-3}$ &5.28$\times 10^{-3}$  &-4.82$\times 10^{-3}$ & -4.95$\times 10^{-3}$ & 0.0945 & 0.0777\\
&S.E.&2.66$\times 10^{-3}$&4.56$\times 10^{-3}$&4.66$\times 10^{-3}$ & 3.13$\times 10^{-3}$ & 0.0122 & 3.82$\times 10^{-3}$\\
&p-value&0.414&0.247&0.301 & 0.114 & 9.30$\times 10^{-15}$ & 0\\
\\\multirow{3}{*}{SW-ps}& Estimate	& 1.09$\times 10^{-3}$ &5.05$\times 10^{-3}$  &-4.62$\times 10^{-3}$ & -4.97$\times 10^{-3}$ & 0.108 & 0.0775\\
&S.E.&2.54$\times 10^{-3}$&4.59$\times 10^{-3}$&4.68$\times 10^{-3}$ & 3.14$\times 10^{-3}$ & 0.0109 & 3.79$\times 10^{-3}$\\
&p-value&0.667&0.271&0.324 & 0.114 & 0 & 0\\
\\\multirow{3}{*}{EL-$\hat{\pi}\hat{\theta}$-5}& Estimate	& 2.57$\times 10^{-5}$ &5.49$\times 10^{-3}$  &-5.00$\times 10^{-3}$ & -4.80$\times 10^{-3}$ & 0.0954 & 0.0777\\
&S.E.&1.45$\times 10^{-3}$&4.57$\times 10^{-3}$&4.69$\times 10^{-3}$ & 3.18$\times 10^{-3}$ & 5.95$\times 10^{-3}$ & 1.76$\times 10^{-3}$\\
&p-value&0.986&0.230&0.286 & 0.131 &0 & 0\\
\\\multirow{3}{*}{EL-$\hat{\pi}\hat{\theta}$-5-ps}& Estimate	& -1.47$\times 10^{-4}$ &5.24$\times 10^{-3}$  &-5.15$\times 10^{-3}$ & -4.99$\times 10^{-3}$ & 0.0952 & 0.0777\\
&S.E.&1.45$\times 10^{-3}$&4.45$\times 10^{-3}$&4.56$\times 10^{-3}$ & 3.10$\times 10^{-3}$ & 5.94$\times 10^{-3}$ & 1.75$\times 10^{-3}$\\
&p-value&0.919&0.240&0.258 & 0.108 & 0 & 0\\
}
\end{scriptsize}

\section{Discussion}

In this article we considered Phase 2 sampling based on $Y_i$ and $\boldsymbol{X}_i$, which has been extensively used (e.g., \cite{barnett2013detecting,lin2013quantitative,derkach2015score,espin2018two}). The EL estimators we proposed make use of the information in Phase 1 data by a model $f(Y|\boldsymbol{X};\boldsymbol{\theta})$ and a selection model $\pi(Y,\boldsymbol{X};\boldsymbol{\alpha})$ for $P(R=1|Y,\boldsymbol{X})$. The model $\pi(Y,\boldsymbol{X};\boldsymbol{\alpha})$ is also essential for CML estimators. However, CML is based on the conditional independence (given $\boldsymbol{X}_i$, $\boldsymbol{Z}_i$, $R_i=1$) of the $Y_i$ and is not applicable when the $R_i$ are not mutually independent, such as residual- or rank-based sampling (e.g \cite{lin2013quantitative} and \cite{derkach2015score}).

Our recommendation of $\widehat{\boldsymbol{\beta}}_{\text{\ensuremath{\widehat{\pi}\widehat{\theta}}-5}}$
is based on its high estimation efficiency and circumvention of potential
multicollinearity of constraints. Its numerical performance is excellent
in our simulation studies. However, the numerical performance may deteriorate when the parameters $\boldsymbol{\alpha}$ and/or
$\boldsymbol{\theta}$ have a large dimension, leading to many constraints
in defining $\widehat{\boldsymbol{\beta}}_{\text{\ensuremath{\widehat{\pi}\widehat{\theta}}-5}}$.
In this case, estimators similar to $\widehat{\boldsymbol{\beta}}_{\text{\ensuremath{\pi\theta^{\ast}}-1}}$
defined by (\ref{EL0-1}) or $\widehat{\boldsymbol{\beta}}_{\text{\ensuremath{\pi\theta_{0}}-1}}$
defined by (\ref{EL0-1-zero}), but with $\widehat{\boldsymbol{\alpha}}_{\text{MLE}}$
and/or $\widehat{\boldsymbol{\theta}}$ estimated separately beforehand
and plugged in, may have a better numerical performance.

It is seen from the simulation studies the proposed method substantially improves the estimation efficiency for two-phase studies. When the fully available covariate is a surrogate of the expensive covariate, the efficiency gain depends on the correlation between the surrogate and the expensive covariate. Higher correlation leads to more efficiency gain. When the fully available and expensive covaraites are both influential and have little correlation, efficiency gains of our proposed method are mainly on the fully available covariates. This observation agrees well with the existing literature on data integration (e.g. \cite{chatterjee2016constrained,han2019empirical}) and is reasonable since the Phase 1 model provides information about the association between $Y$ and $\boldsymbol{X}$. The proposed method is hence advantageous when an inexpensive surrogate $\boldsymbol{X}$ of a Phase 2 covariate $\boldsymbol{Z}$ is available for all Phase 1 individuals, or when the interest is in the association between the response and the covariates $\boldsymbol{X}$, and one wants to adjust for some expensive or hard-to-collect covariates $\boldsymbol{Z}$, e.g., the scenario we study in the real data analysis. By making use of the available Phase 1 data that do not contain $\boldsymbol{Z}$, the proposed method produces substantial efficiency gains compared to methods that use Phase 2 data alone either on estimating effects of $\boldsymbol{Z}$ when $\boldsymbol{X}$ is not included in the outcome model of interest; or on $\boldsymbol{X}$ when both covariates are included. For example, in Table 3, our recommended EL estimator yields similar standard errors for $\boldsymbol{X}$ compared to the MLE using the full data, and that is achieved by selecting only 20\% of the Phase 1 data into Phase 2 to collect $\boldsymbol{Z}$. In many studies such an efficiency improvement could substantially reduce the study cost by collecting $\boldsymbol{Z}$ only from a subset of the sample without compromising the conclusion on the effect of primary interest. 

The proposed method is connected to some recent literature on certain
data integration problems (e.g.,\cite{chatterjee2016constrained,huang2016efficient,han2019empirical,han2022integrating}), where an external big data source
provides some summary information that can be used to improve the
efficiency for internal study model fitting. Our method treats the
Phase 1 data as an external data source whose information is used to improve
the efficiency for estimation based on Phase 2 data. There are two major differences between their setup and ours. First, in
two-phase samples, the Phase 2 sample is a subset of the Phase 1 sample, while in
most prior literature, the “internal" and “external" studies are two independent
samples. Second, prior methods often
assume only “summary-level" information is obtained from the external
dataset, while in two-phase studies, individual-level data of every individual
in Phase 1 is known. Therefore, we may incorporate different auxiliary information, e.g., modeling $P(R=1|Y,\boldsymbol{X})=\pi(Y,\boldsymbol{X}; \boldsymbol{\alpha})$ and estimating $h(Y,\boldsymbol{X};\boldsymbol{\theta})$. Unlike prior literatures in which the likelihood for the ``internal study" that is directly applicable, for ODS, due to the selection probability, the internal likelihood needs to be adjusted, which we used conditional likelihood. In the simplified case that the selection probability is known, and Phase 1 sample is so large that the uncertainty in estimating $\boldsymbol{\theta}$ can be ignored, if we treat the conditional likelihood as the internal likelihood in prior literature, our proposed estimator  $\widehat{\boldsymbol{\beta}}_{\text{\ensuremath{\pi\theta^{\ast}}-1}}$ has the same asysmptotic efficiency as \cite{chatterjee2016constrained} (see Section \ref{proof1}). However as we explained before, when not ignoring the uncertainty in estimating $\boldsymbol{\theta}$, the asymptotic efficiencies are not directly comparable. %The method can be naturally generalized when the internal study involves an ODS study hence conditional likelihood is needed, and there are external studies that provide useful summary-level information. 

\section{Appendix}
\subsection{Asymptotic Distribution of  $\sqrt{n}(\hat{\boldsymbol{\beta}}_{\pi\theta^{\ast}-1}-\boldsymbol{\beta}_0)$}
\label{proof1}
As the $\boldsymbol{\theta}^\ast$ is fixed through the optimization, we omit $\boldsymbol{\theta}$ in the function $\boldsymbol{u}(\boldsymbol{\beta},\boldsymbol{\theta})$. The Lagrangian of the corresponding optimization is 
\begin{align*}
  \mathcal{L}=\sum_{i=1}^n(\log f_{c,i}+\log p_i)+n\boldsymbol{\lambda}^T\sum_{i=1}^np_iR_i\bfu_i(\boldsymbol{\beta})+\mu(\sum_{i=1}^np_i-1),
\end{align*}

At the solution $\widehat{\boldsymbol{\beta}}_{\text{\ensuremath{\pi\theta^{\ast}}-1}}$ and $\hat{p}_i$ we have $\partial\mathcal{L}/\partial p_i=0$ and $\partial\mathcal{L}/\partial{\boldsymbol{\beta}}=0$, which yields 
\[\hat{p}_i=1/[n\{1-\widehat{\boldsymbol{\lambda}}^TR_i\bfu_i(\widehat{\boldsymbol{\beta}}_{\text{\ensuremath{\pi\theta^{\ast}}-1}})\}]\] and
\[{\bf0}=\sum_{i=1}^n R_i\bfs_{c,\beta,i}(\widehat{\boldsymbol{\beta}}_{\text{\ensuremath{\pi\theta^{\ast}}-1}})+ \sum_{i=1}^n\hat{p}_iR_i\partial \boldsymbol{u}_i(\widehat{\boldsymbol{\beta}}_{\text{\ensuremath{\pi\theta^{\ast}}-1}})/\partial\boldsymbol{\beta}^T\widehat{\boldsymbol{\lambda}},\]
\[{\bf0}=\sum_{i=1}^n\hat{p}_iR_i\boldsymbol{u}_i(\widehat{\boldsymbol{\beta}}_{\text{\ensuremath{\pi\theta^{\ast}}-1}}).\]

Applying the mean value theorem to the last two equations around $\boldsymbol{\beta}_0$ and $\bf0$, we have 
\begin{align*}
    {\bf0}=&\left(\begin{array}{c}
         1/n\sum_{i=1}^n R_i\bfs_{c,\beta,i}(\boldsymbol{\beta}_0)  \\
         1/n\sum_{i=1}^n R_i\bfu_i(\boldsymbol{\beta}_0)
    \end{array}\right)\\
    +&\left(\begin{array}{cc}
    1/n\sum_{i=1}^n \frac{R_i\partial \bfs_{c,\beta,i}(\overline{\boldsymbol{\beta}})}{\partial \boldsymbol{\beta}}& 1/n\sum_{i=1}^n \frac{R_i\partial \bfu_i (\widehat{\boldsymbol{\beta}}_{\text{\ensuremath{\pi\theta^{\ast}}-1}})/\partial \boldsymbol{\beta}^T}{1-\widehat{\boldsymbol{\lambda}}^TR_i\bfu_i(\widehat{\boldsymbol{\beta}}_{\text{\ensuremath{\pi\theta^{\ast}}-1}})}\\
    1/n\sum_{i=1}^n \frac{\partial R_i\bfu_i(\overline{\boldsymbol{\beta}})/\partial \boldsymbol{\beta}}{1-\overline{\boldsymbol{\lambda}}^TR_i\bfu_i(\widehat{\boldsymbol{\beta}}_{\text{\ensuremath{\pi\theta^{\ast}}-1}})}&1/n\sum_{i=1}^n \frac{ R_i\bfu_i(\boldsymbol{\beta}_0)\bfu_i(\widehat{\boldsymbol{\beta}}_{\text{\ensuremath{\pi\theta^{\ast}}-1}})^T}{\{1-\overline{\boldsymbol{\lambda}}^T\bfu_i(\widehat{\boldsymbol{\beta}}_{\text{\ensuremath{\pi\theta^{\ast}}-1}})\}^2}
 %        1/m\sum_{i=1}^m \frac{\partial s_i(\boldsymbol{\beta}_0)}{\partial \boldsymbol{beta}}&1/m\sum_{i=1}^m \frac{\partial u_i(\boldsymbol{\beta}_0)/\partial \boldsymbol{beta}}}{1-\hat{\lambda}^{T}u_i(\hat{\boldsymbol{\beta}}_\text{SEL1})}   \\
 %        1/m\sum_{i=1}^m u_i(\boldsymbol{\beta}_0&
    \end{array}\right)\left(\begin{array}{c}
        \widehat{\boldsymbol{\beta}}_{\text{\ensuremath{\pi\theta^{\ast}}-1}}-\boldsymbol{\beta}_0  \\
         \widehat{\boldsymbol{\lambda}}
    \end{array}\right)\\
    +&o_p(\|(\widehat{\boldsymbol{\beta}}_{\text{\ensuremath{\pi\theta^{\ast}}-1}}^T-\boldsymbol{\beta}_0^T,\ \widehat{\boldsymbol{\lambda}}^T)\|)\\
    :=&\boldsymbol{v}_n+\bfG_n\left(\begin{array}{c}
        \widehat{\boldsymbol{\beta}}_{\text{\ensuremath{\pi\theta^{\ast}}-1}}-\boldsymbol{\beta}_0  \\
         \widehat{\boldsymbol{\lambda}}
    \end{array}\right) +o_p(\|(\widehat{\boldsymbol{\beta}}_{\text{\ensuremath{\pi\theta^{\ast}}-1}}^T-\boldsymbol{\beta}_0^T,\ \widehat{\boldsymbol{\lambda}}^T)\|)
\end{align*}
for some $\overline{\boldsymbol{\beta}}$ between $\boldsymbol{\beta}_0$  and $\widehat{\boldsymbol{\beta}}_{\text{\ensuremath{\pi\theta^{\ast}}-1}}$ and some $\overline{\boldsymbol{\lambda}}$ between $\boldsymbol{0}$  and $\widehat{\boldsymbol{\lambda}}$.
Then we have 
\begin{align*}
    \sqrt{n}\boldsymbol{v}_n \xrightarrow{d}N({\bf0},\boldsymbol{\Sigma})
\end{align*}
where 
\begin{align*}
   \boldsymbol{\Sigma}=\left(\begin{array}{cc}
    \boldsymbol{S}  &  \bf0\\
    \bf0    & \boldsymbol{\Omega}
   \end{array}
   \right),
\end{align*}
with $\bfS=E\{R\boldsymbol{s}_{c,\beta}(\boldsymbol{\beta}_0)\boldsymbol{s}_{c,\beta}(\boldsymbol{\beta}_0)^T\}$, $\boldsymbol{\Omega}=E\{R\bfu(\boldsymbol{\beta}_0)\bfu(\boldsymbol{\beta}_0)^T\}$. The off-diagonal block is zero as 
\begin{align*}
    E\{R\bfs_{c,\beta}(\boldsymbol{\beta}_0)\bfu(\boldsymbol{\beta}_0)^T\}&=P(R=1)E\{\bfs_{c,\beta}(Y,\bfX,\bfZ;\boldsymbol{\beta}_0)\bfu(\bfX,\bfZ;\boldsymbol{\beta}_0)|R=1\}\\
    &=P(R=1)E[E\{\bfs_{c,\beta}(Y,\bfX,\bfZ;\boldsymbol{\beta}_0)\bfu(\bfX,\bfZ;\boldsymbol{\beta}_0)|\bfX,\bfZ,R=1\}]\\
    &=P(R=1)E[E\{\bfs_{c,\beta}(Y|\bfX,\bfZ,R=1;\boldsymbol{\beta}_0)\}\bfu(\bfX,\bfZ|R=1;\boldsymbol{\beta}_0)]
\end{align*}
and we know $E\{\bfs_{c,\beta}(Y|\bfX,\bfZ,R=1;\boldsymbol{\beta}_0)\}=\bf0$.

Denoting $\bfJ=E\{R\partial \bfu(\boldsymbol{\beta}_0)/\partial \boldsymbol{\beta}^T\}$, we have 
\begin{align*}
    \bfG_n\xrightarrow{p}\bfG:=\left(\begin{array}{cc}
    E\{R\frac{\partial \bfs_{c,\beta}(\boldsymbol{\beta}_0)}{\partial \boldsymbol{\beta}}\}  &  E\{R\frac{\partial \bfu(\boldsymbol{\beta}_0)}{\partial \boldsymbol{\beta}^T}\}\\
    E\{R\frac{\partial \bfu(\boldsymbol{\beta}_0)}{\partial \boldsymbol{\beta}}\}    & E\{R\bfu(\boldsymbol{\beta}_0)\bfu(\boldsymbol{\beta}_0)^T\}
   \end{array}
   \right)=\left(\begin{array}{cc}
    -\bfS  &  \bfJ\\
    \bfJ^T    &  \boldsymbol{\Omega}
   \end{array}
   \right),
\end{align*}
thus 
\begin{align*}
    \sqrt{n}\left(\begin{array}{c}
        \widehat{\boldsymbol{\beta}}_{\pi\theta^{\ast}-1}-\boldsymbol{\beta}_0  \\
         \widehat{\boldlambda}
    \end{array}\right)\xrightarrow{d}N\left({\bf0}, \bfG^{-1}\boldsymbol{\Sigma} \bfG^{-T}\right).
\end{align*}
and $\ACov\{ \sqrt{n}(\widehat{\boldsymbol{\beta}}_{\pi\theta^\ast-1}-\boldsymbol{\beta}_0)\}
$ equals the upper left block of $\bfG^{-1}\boldsymbol{\Sigma} \bfG^{-T}$, which is $\boldsymbol{\Sigma}_{\pi\theta^{\ast}}=(\bfS+\bfJ\boldsymbol{\Omega}^{-1} \bfJ^T)^{-1}$.

In \cite{chatterjee2016constrained}, the constrained maximum likelihood estimator has asymptotic variance $(\bfS+\boldsymbol{C}\boldsymbol{\Sigma}^{-1}\boldsymbol{C})^{-1}$, where $\boldsymbol{C}=E\{\bfs_{c,\beta}(Y|\bfX,\bfZ,R=1;\boldsymbol{\beta}_0)\bfh^T(\bfX,\bfZ|R=1;\boldsymbol{\beta}_0)\}$. Note that 
\begin{align*}
   \bfJ=&E\{R\partial \bfu(\boldsymbol{\beta}_0)/\partial \boldsymbol{\beta}^T\}=E\left\{R\frac{\partial}{\partial \boldsymbol{\beta}^T}\int \frac{\bfh}{\pi}f_cdY\right\}\\
   =&E\left\{\int \pi \frac{\boldsymbol{h}}{\pi}\frac{\partial f_c}{\partial \boldsymbol{\beta}^T}dY\right\}=E\left\{\int \boldsymbol{h}\frac{1}{f_c}\frac{\partial f_c}{\partial \boldsymbol{\beta}^T}f_cdY\right\}\\
   =&E\{\boldsymbol{s}_{c,\boldsymbol{\beta}}\bfh^T\}=\boldsymbol{C},
\end{align*}
thus $(\bfS+\boldsymbol{C}\boldsymbol{\Sigma}^{-1}\boldsymbol{C})^{-1}=\boldsymbol{\Sigma}_{\pi\theta^{\ast}}$.

\subsection{Equivalence of the $\widehat{\boldsymbol{\beta}}_{\pi\widehat{\theta}-1}$ and $\widehat{\boldsymbol{\beta}}_{\pi\widehat{\theta}-2}$ estimator}
Similarly as in Section \ref{proof1}, we note that $\widehat{\boldsymbol{\beta}}_{\pi\widehat{\theta}-1}$ and the Lagrange multiplier $\widehat{\boldsymbol{\lambda}}$ satisfy
\begin{align*}
    \ \sum_{i=1}^n\left\{R_i\bfs_{c,\beta}(\widehat{\boldsymbol{\beta}}_{\pi\widehat{\theta}-1})+\frac{R_i\partial \bfu_i(\widehat{\boldsymbol{\beta}}_{\pi\widehat{\theta}-1},\widehat{\boldsymbol{\theta}})/\partial \boldsymbol{\beta}^T}{1-\widehat{\boldsymbol{\lambda}}^T\bfu_i(\widehat{\boldsymbol{\beta}}_{\pi\widehat{\theta}-1},\widehat{\boldsymbol{\theta}})}\widehat{\boldsymbol{\lambda}}\right\}&=\bf0\\
\ \ \ \  \sum_{i=1}^n\left\{\frac{R_i\bfu_i(\widehat{\boldsymbol{\beta}}_{\pi\widehat{\theta}-1},\widehat{\boldsymbol{\theta}})}{1-\widehat{\boldsymbol{\lambda}}^T\bfu_i(\widehat{\boldsymbol{\beta}}_{\pi\widehat{\theta}-1},\widehat{\boldsymbol{\theta}})}\right\}&=\bf0,\\
     \mbox{and}\ \ \ \ \ \ \ \ \ \ \
     \sum_{i=1}^n\{\bfh(Y_i,\bfX_i;\widehat{\boldsymbol{\theta}})\}&=\bf0.
\end{align*}
By a first-order Taylor expansion, 
\begin{align*}
  \nonumber {\bf0}&= \frac{1}{n}\sum_{i=1}^n\left(\begin{array}{c}
         R_i\bfs_{c,\beta,i}(\boldsymbol{\beta}_0)  \\
         R_i\bfu_i(\boldsymbol{\beta}_0,\boldsymbol{\theta}^\ast)\\
         \bfh(\boldsymbol{\theta}^\ast)
    \end{array}\right)+\frac{1}{n}\sum_{i=1}^n\left(\begin{array}{ccc}
          \frac{R_i\partial \bfs_{c,\beta,i}(\boldsymbol{\beta}_0)}{\partial\boldsymbol{\beta}}&\frac{\partial R_i\bfu_i/\partial \boldsymbol{\beta}^T}{1-\widehat{\boldsymbol{\lambda}}^T\bfu_i}&{\bf0}\\
         \frac{R_i\partial \bfu_i/\partial \boldsymbol{\beta}}{1-\widehat{\boldsymbol{\lambda}}^T\bfu_i}&\frac{ R_i\bfu_i\bfu_i^T}{(1-\widehat{\boldsymbol{\lambda}}^T\bfu_i)^2}&\frac{R_i\partial \bfu_i/\partial \boldsymbol{\theta}}{1-\widehat{\boldsymbol{\lambda}}^T\bfu_i}\\
         \bf0&\bf0&\frac{\partial \bfh(\boldsymbol{\theta}^\ast)}{\partial \boldsymbol{\theta}}
    \end{array}\right)\\&\cdot \left(\begin{array}{c}
         \widehat{\boldsymbol{\beta}}_{\pi\widehat{\theta}-1}-\boldsymbol{\beta}_0  \\
         \widehat{\boldsymbol{\lambda}}\\
         \widehat{\boldsymbol{\theta}}-\boldsymbol{\theta}^\ast
    \end{array}\right)+o_p(1)\\
    &:= \bfv_n+\bfG_n\left(\begin{array}{c}
         \widehat{\boldsymbol{\beta}}_{\pi\widehat{\theta}-1}-\boldsymbol{\beta}_0  \\
         \widehat{\boldsymbol{\lambda}}\\
        \widehat{\boldsymbol{\theta}}-\boldsymbol{\theta}^\ast
    \end{array}\right)+o_p(1),
\end{align*}
and by CLT and law of large numbers,
\begin{align*}
    \sqrt{n}\bfv_n \xrightarrow{d}N({\bf0},\bfSigma)
\end{align*}
where 
\begin{align*}
   \bfSigma=\left(\begin{array}{ccc}
    \bfS  &  \bf0 & \bfU^T\\
    \bf0    & \bfOmega & \bfV^T\\
    \bfU & \bfV & \bfW
   \end{array}
   \right),
\end{align*}
where $\bfW=E(\bfh(\boldsymbol{\theta}^*)\bfh(\boldsymbol{\theta}^*)^T)$;
 $\bfG_n$ converges in probability to constant matrix
$$\bfG:=\left(\begin{array}{ccc}
     -\bfS&\bfJ & \bf0\\
     \bfJ^T& \bfOmega& -\bfH \\
     \bf0&\bf0&-\tilde{\bfW}
\end{array}\right)=\left(\begin{array}{ccc}
     E\{R\partial \bfs_{c,\beta}/\partial \boldsymbol{\beta}\}&  E\{R\partial \bfu/\partial \boldsymbol{\beta}^T\} & \bf0\\
      E\{R\partial \bfu/\partial \boldsymbol{\beta}\}& E\{R\bfu\bfu^T\} &E\{R\partial \bfu/\partial \boldsymbol{\theta}\}\\
      \bf0&\bf0&E\{\partial \bfh/\partial \boldsymbol{\theta}\}
\end{array}\right)
,$$
where each function is evaluated at $\boldsymbol{\beta}_0$ and $\boldsymbol{\theta}^*$. Thus 
\begin{align*}
    \sqrt{n}\left(\begin{array}{c}
        \widehat{\boldsymbol{\beta}}_{\pi\widehat{\theta}-1}-\boldsymbol{\beta}_0  \\
         \widehat{\boldsymbol{\lambda}}\\
          \widehat{\boldsymbol{\theta}}-\boldsymbol{\theta}^\ast
    \end{array}\right)\xrightarrow{d}N\left({\bf0}, \bfG^{-1}\bfSigma \bfG^{-T}\right).
\end{align*}
It is to be noted that 
\begin{align*}
    \bfu(\bfX,\bfZ;\boldtheta)&=E\left\{\frac{\bfh(Y,\bfX; \boldtheta)}{\pi(Y,\bfX)}\vert \bfX,\bfZ,R=1\right\}\\
    &=\frac{\int \bfh(Y,\bfX; \boldtheta)f(Y|\bfX,\bfZ)dY}{\int f(Y|\bfX,\bfZ)\pi(Y,\bfX)dY}
\end{align*}
thus
\begin{align*}
   &E\left\{\frac{R\partial \bfu(\bfX,\bfZ;\boldtheta)}{\partial \boldsymbol{\theta}}\right\}=\int R\frac{\partial\bfu(\bfX,\bfZ;\boldtheta)}{\partial \boldsymbol{\theta}}f(\bfX,\bfZ,R)dRd\bfX d\bfZ\\
   =&\int \frac{\partial\bfu(\bfX,\bfZ;\boldtheta)}{\partial \boldsymbol{\theta}}f(\bfX,\bfZ,R=1)d\bfX d\bfZ\\
    =&\int \frac{\partial\bfu(\bfX,\bfZ;\boldtheta)}{\partial \boldsymbol{\theta}}\left\{\int f(Y,\bfX,\bfZ)\pi(Y,\bfX)dY\right\}d\bfX d\bfZ\\
   =&\int\left[\frac{\partial }{\partial \boldsymbol{\theta}}E\left\{\frac{\bfh(Y,\bfX; \boldtheta)}{\pi(Y,\bfX)}\vert \bfX,\bfZ,R=1\right\}\right]\left\{\int f(Y\vert\bfX,\bfZ)g(\bfX,\bfZ)\pi(Y,\bfX)dY\right\}d\bfX d\bfZ\\
   =&\int \left\{\frac{\int\partial \bfh(Y,\bfX; \boldtheta)/\partial \boldsymbol{\theta}f(Y|\bfX,\bfZ)dY}{\int f(Y|\bfX,\bfZ)\pi(Y,\bfX)dY}\right\}\left\{\int f(Y|\bfX,\bfZ)\pi(Y,\bfX)dY\right\}g(\bfX,\bfZ)d\bfX d\bfZ\\
   =&\int\left\{\int\frac{\partial \bfh(Y,\bfX; \boldtheta)}{\partial \boldsymbol{\theta}}f(Y|\bfX,\bfZ)dY\right\}g(\bfX,\bfZ)d\bfX d\bfZ=E\left\{\frac{\partial \bfh(Y,\bfX; \boldtheta)}{\partial \boldsymbol{\theta}}\right\}
\end{align*}
therefore $\bfH=\tilde{\bfW}$. We then have 
\begin{align*}
\bfG^{-1}=&\left(\begin{array}{ccc}
     -\bfS&\bfJ & \bf0\\
     \bfJ^T& \bfOmega&-\bfH \\
     \bf0&\bf0&-\bfH
\end{array}\right)^{-1}=\left(\begin{array}{ccc}
   \bfG^{11}&\bfG^{12}\\
   \bf0& \bfG^{22}
\end{array}\right)
\end{align*}
with \begin{align*}
    \bfG^{11}&=\left(\begin{array}{cc}-\bfS&\bfJ \\
     \bfJ^T& \bfOmega\end{array} \right)^{-1}\\
     \bfG^{12}&=\left(\begin{array}{cc}-\bfS&\bfJ \\
     \bfJ^T& \bfOmega\end{array}\right)^{-1}\left(\begin{array}{c}\bf0 \\
     -\bfH\end{array} \right)\bfH^{-1}\\
    \bfG^{22}&=-\bfH^{-1}.
\end{align*}
%\left(\begin{array}{cc}
%     \left(\begin{array}{cc}-\bfS&\bfJ \\
%     \bfJ^T& \bfOmega\end{array} \right)^{-1}&  \left(\begin{array}{cc}-\bfS&\bfJ \\
%     \bfJ^T& \bfOmega\end{array} \right)^{-1} \left(\begin{array}{c}\bf0 \\
%     -\bfH\end{array} \right)\bfH^{-1}\\
%     \begin{array}{cc}
%     \bf0   \  &\ \ \ \ \bf0
%     \end{array}&-\bfH^{-1}
%\end{array}\right)\\
and thus 
\begin{align*}
\bfG^{-1}=&\left(\begin{array}{ccc}
     -(\bfS+\bfJ\bfOmega^{-1} \bfJ^T)^{-1}\ \ \ \ \  &(\bfS+\bfJ\bfOmega^{-1} \bfJ^T)^{-1}\bfJ\bfOmega^{-1}\ \ \ \ \ \ & -(\bfS+\bfJ\bfOmega^{-1} \bfJ^T)^{-1}\bfJ\bfOmega^{-1}\\
     *& *& * \\
     \bf0& \bf0&*
\end{array}\right)\\
=&(\bfS+\bfJ\bfOmega^{-1} \bfJ^T)^{-1}\left(\begin{array}{ccc}
     -\bfI&\bfJ\bfOmega^{-1}& -\bfJ\bfOmega^{-1}\\
     *& *& * \\
     \bf0&\bf0&*
\end{array}\right).
\end{align*}
Hence the asymptotic covariance of $\sqrt{n}(\widehat{\boldsymbol{\beta}}_{\pi\widehat{\theta}-1}-\boldsymbol{\beta}_0)$, i.e., the top-left block of $\bfG^{-1}\bfSigma\bfG^{-T}$ is  \begin{align*}
    &\ACov\{ \sqrt{n}(\widehat{\boldsymbol{\beta}}_{\pi\widehat{\theta}-1}-\boldsymbol{\beta}_0)\}\\
    =&(\bfS+\bfJ\bfOmega^{-1} \bfJ^T)^{-1}\left(\begin{array}{ccc}
     -\bfI&\bfJ\bfOmega^{-1}& -\bfJ\bfOmega^{-1}\end{array}\right)\left(\begin{array}{ccc}
    \bfS  &  \bf0 & \bfU^T\\
    \bf0    & \bfOmega & \bfV^T\\
    \bfU & \bfV & \bfW
   \end{array}
   \right)\left(\begin{array}{c}
     -\bfI\\ \bfOmega^{-1}\bfJ^T\\-\bfOmega^{-1}\bfJ^T\end{array}\right)(\bfS+\bfJ\bfOmega^{-1} \bfJ^T)^{-1}\\
   %  =&(S+J^T\Omega^{-1} J)^{-1}\{S+J^T\Omega^{-1}J-J^T\Omega^{-1}V\Omega^{-1}J-J^T\Omega^{-1}V^{T}\Omega^{-1}J+J^T\Omega^{-1}W\Omega^{-1}J\}(S+J^T\Omega^{-1} J)^{-1}\\
     =&(\boldsymbol{S}+\boldsymbol{J}\boldsymbol{\Omega}^{-1}\boldsymbol{J}^{T})^{-1}\left\{ \boldsymbol{S}+\boldsymbol{J}\boldsymbol{\Omega}^{-1}\boldsymbol{U}+\boldsymbol{U}^{T}\boldsymbol{\Omega}^{-1}\boldsymbol{J}^{T}+\boldsymbol{J}\boldsymbol{\Omega}^{-1}(\boldsymbol{\Omega}-\boldsymbol{V}-\boldsymbol{V}^{T}+\boldsymbol{W})\boldsymbol{\Omega}^{-1}\boldsymbol{J}^{T}\right\}\\ \cdot&(\boldsymbol{S}+\boldsymbol{J}\boldsymbol{\Omega}^{-1}\boldsymbol{J}^{T})^{-1}.
\end{align*}
%There is no direct comparison between $\ACov\{ \sqrt{n}(\widehat{\boldsymbol{\beta}}_\text{EL0-1}-\boldsymbol{\beta}_0)\}$ and $\ACov\{ \sqrt{n}(\widehat{\boldsymbol{\beta}}_\text{EL0-2}-\boldsymbol{\beta}_0)\}$. However, w
%We now show the equivalence of $\widehat{\boldsymbol{\beta}}_{\pi\widehat{\theta}-1}$ and $\widehat{\boldsymbol{\beta}}_{\pi\widehat{\theta}-2}$. 
The $\widehat{\boldsymbol{\beta}}_{\pi\widehat{\theta}-2}$ estimator is defined through \begin{align*}
 &\max_{\boldsymbol{\beta},p_1,...,p_n}\prod_{i=1}^n p_i \ \mbox{ subject to }p_i\ge 0,\ \sum_{i=1}^np_i=1,\  \\
    \mbox{and} \ & \sum_{i=1}^np_i\left(\begin{array}{c}R_i\bfu(\bfX_i,\bfZ_i,\boldsymbol{\beta}, \widehat{\boldsymbol{\theta}}
    )\\
    R_i\bfs_{c,\beta}(Y_i,\bfX_i,\bfZ_i;\boldsymbol{\beta})\end{array}\right)=\bf0.
    \label{beta_EL2}
\end{align*}
By Lagrange multipliers method as in \cite{qin1994empirical}, one can write the profile likelihood as 
\[l(\boldsymbol{\beta},\boldsymbol{\lambda},\boldsymbol{\theta})=-\sum_{i=1}^n\log\{1-\boldsymbol{\lambda}^T\bfg_i(\boldsymbol{\beta},\boldsymbol{\theta})\}\]
where $\bfg_i(\boldsymbol{\beta},\boldsymbol{\theta})=(R_i\bfs_{c,\beta}(Y_i,\bfX_i,\bfZ_i,\boldsymbol{\beta})^T,R_i\bfu(\bfX_i,\bfZ_i,\boldsymbol{\beta},{\boldsymbol{\theta}})^T)^T.$ Write 
\begin{align*}
    \bfQ_{1n}(\boldsymbol{\beta},\boldsymbol{\lambda},\boldsymbol{\theta})&=-\frac{1}{n}\frac{\partial l}{\partial \boldsymbol{\lambda}}=\frac{1}{n}\sum_{i=1}^n\frac{ \bfg_i(\boldsymbol{\beta},\boldsymbol{\theta})}{1-\boldsymbol{\lambda}^T\bfg_i(\boldsymbol{\beta},\boldsymbol{\theta})},\\
      \bfQ_{2n}(\boldsymbol{\beta},\boldsymbol{\lambda},\boldsymbol{\theta})&=-\frac{1}{n}\frac{\partial l}{\partial \boldsymbol{\beta}}=\frac{1}{n}\sum_{i=1}^n\frac{\boldsymbol{\lambda}^T\partial \bfg_i(\boldsymbol{\beta},\boldsymbol{\theta})/\partial \boldsymbol{\beta} }{1-\boldsymbol{\lambda}^T\bfg_i(\boldsymbol{\beta},\boldsymbol{\theta})},\\
      \bfQ_{3n}(\boldsymbol{\theta})&=\frac{1}{n}\sum_{i=1}^n \bfh_i(\boldsymbol{\theta}).
\end{align*}
Let $\widehat{\boldsymbol{\lambda}}$ denote the Lagrange multiplier that satisfies the constraint \[\sum_{i=1}^n\frac{ \bfg_i(\boldsymbol{\beta},\boldsymbol{\theta})}{1-\boldsymbol{\lambda}^T\bfg_i(\boldsymbol{\beta},\boldsymbol{\theta})}=\bf0,\]
then by Z-estimator theory, $\widehat{\boldsymbol{\beta}}_{\pi\widehat{\theta}-2}$ is a consistent estimator; $\widehat{\boldsymbol{\theta}}$ converges in probability to $\boldtheta^\ast$; and by general empirical likelihood thoery, $\widehat{\boldsymbol{\lambda}}=O_p(n^{-1/2})$. An application of a first-order Taylor expansion around $(\boldsymbol{\beta}_0,\bf0,\boldsymbol{\theta}^\ast)$ yields
\begin{align*}
{\bf0}&= \bfQ_{1n}(\boldsymbol{\beta}_0,{\bf0},\boldsymbol{\theta}^\ast)+\frac{\partial \bfQ_{1n}(\boldsymbol{\beta}_0,\bf0,\boldsymbol{\theta}^\ast)}{\partial \boldsymbol{\beta}^T}(\widehat{\boldsymbol{\beta}}_{\pi\widehat{\theta}-2}-\boldsymbol{\beta}_0) +\frac{\partial \bfQ_{1n}(\boldsymbol{\beta}_0,\bf0,\boldsymbol{\theta}^\ast)}{\partial \boldsymbol{\lambda}^T}\widehat{\boldsymbol{\lambda}}\\&+\frac{\partial \bfQ_{1n}(\boldsymbol{\beta}_0,\bf0,\boldsymbol{\theta}^\ast)}{\partial \boldsymbol{\theta}^T}(\widehat{\boldsymbol{\theta}}-\boldsymbol{\theta}^\ast)+o_p(n^{-1/2}); \\
{\bf0}&= %\bfQ_{2n}(\boldsymbol{\beta}_0,{\bf0},\boldsymbol{\theta}^\ast)+\frac{\partial \bfQ_{2n}(\boldsymbol{\beta}_0,\bf0,\boldsymbol{\theta}^\ast)}{\partial \boldsymbol{\beta}^T}(\widehat{\boldsymbol{\beta}}_{\pi\widehat{\theta}-2}-\boldsymbol{\beta}_0) +\frac{\partial \bfQ_{2n}(\boldsymbol{\beta}_0,\bf0,\boldsymbol{\theta}^\ast)}{\partial \boldsymbol{\lambda}^T}\widehat{\boldsymbol{\lambda}}\\&+\frac{\partial \bfQ_{2n}(\boldsymbol{\beta}_0,\bf0,\boldsymbol{\theta}^\ast)}{\partial \boldsymbol{\theta}^T}(\widehat{\boldsymbol{\theta}}-\boldsymbol{\theta}^\ast)+O_p(n^{-1}) \\
\widetilde{\bfS}(\boldbeta_0,{\bf0},\boldtheta^\ast)\widehat{\boldsymbol{\lambda}}+o_p(n^{-1/2}) \mbox{, where } \widetilde{\bfS}(\boldbeta,\boldlambda,\boldtheta)=\frac{1}{n}\sum_{i=1}^n\frac{\partial \bfg_i(\boldbeta,\boldtheta)/\partial \boldbeta}{1-\boldlambda^T\bfg_i({\boldbeta},{\boldtheta})};\\
{\bf0}&=\bfQ_{3n}(\boldsymbol{\beta}_0,{\bf0},\boldsymbol{\theta}^\ast)+\frac{ \partial \bfQ_{3n}(\boldsymbol{\beta}_0,{\bf0},\boldsymbol{\theta}^\ast)}{\partial \boldsymbol{\theta}^T}(\widehat{\boldsymbol{\theta}}-\boldsymbol{\theta}^\ast)+o_p(n^{-1/2}). 
\end{align*}
From the last equation, 
\begin{align*}
   \widehat{\boldsymbol{\theta}}-\boldsymbol{\theta}^\ast=&- \left\{\frac{ \partial \bfQ_{3n}(\boldsymbol{\beta}_0,{\bf0},\boldsymbol{\theta}^\ast)}{\partial \boldsymbol{\theta}^T}\right\}^{-1}\bfQ_{3n}(\boldsymbol{\beta}_0,{\bf0},\boldsymbol{\theta}^\ast)+o_p(n^{-1/2})\\
   =&\bfH^{-1}\bfQ_{3n}+o_p(n^{-1/2}),
\end{align*}
thus 
\begin{align}
  \nonumber {\bf0}&= \left(\begin{array}{c}
         \bfQ_{1n}(\boldsymbol{\beta}_0,\bf0,\boldsymbol{\theta}^\ast)  \\
         \bf0\\
    \end{array}\right)+\left(\begin{array}{cc}
          \frac{\partial \bfQ_{1n}(\boldsymbol{\beta}_0,{\bf0},\boldsymbol{\theta}^\ast)}{\partial\boldsymbol{\lambda}^T}&\frac{\partial \bfQ_{1n}(\boldsymbol{\beta}_0,{\bf0},\boldsymbol{\theta}^\ast)}{\partial\boldsymbol{\beta}^T}\\
         \widetilde{\bfS}(\boldbeta_0,\boldsymbol{0},\boldtheta^\ast)&\bf0
    \end{array}\right)\cdot \left(\begin{array}{c}
      \widehat{\boldsymbol{\lambda}}\\
         \widehat{\boldsymbol{\beta}}_{\pi\widehat{\theta}-2}-\boldsymbol{\beta}_0  
       %  \widehat{\boldsymbol{\theta}}_\text{MLE}-\boldsymbol{\theta}_0
    \end{array}\right)\\
    &+\left(\begin{array}{c}\frac{\partial \bfQ_{1n}(\boldsymbol{\beta}_0,{\bf0},\boldsymbol{\theta}^\ast)}{\partial\boldsymbol{\theta}^T}(\bfH^{-1}\bfQ_{3n})\\
   \bf0\end{array}\right)+o_p(n^{-1/2}).
    \label{taylor2}
\end{align}
Noting that \begin{align*}
     \frac{\partial \bfQ_{1n}(\boldsymbol{\beta}_0,{\bf0},\boldsymbol{\theta}^\ast)}{\partial\boldsymbol{\beta}^T}=&\frac{1}{n}\sum_{i=1}^n\frac{ \partial \bfg_i(\boldsymbol{\beta}_0,\boldsymbol{\theta}^\ast)/\partial \boldsymbol{\beta}^T}{1-\boldsymbol{0}^T\bfg_i(\boldsymbol{\beta}_0,\boldsymbol{\theta}^\ast)}+o_p(n^{-1/2})\xrightarrow{p}\left(\begin{array}{c}
      -\bfS     \\
      \bfJ^T   
     \end{array}\right),\\
          \frac{\partial \bfQ_{1n}(\boldsymbol{\beta}_0,{\bf0},\boldsymbol{\theta}^\ast)}{\partial\boldsymbol{\lambda}^T}=&\frac{1}{n}\sum_{i=1}^n\frac{ -\bfg_i(\boldsymbol{\beta}_0,\boldsymbol{\theta}^\ast)\bfg_i(\boldsymbol{\beta}_0,\boldsymbol{\theta}^\ast)^{^T}}{(1-\boldsymbol{0}^T\bfg_i(\boldsymbol{\beta}_0,\boldsymbol{\theta}^\ast))^2}+o_p(n^{-1/2})\xrightarrow{p}\left(\begin{array}{cc}
      -\bfS  & \bf0   \\
     \bf0 & -\bfOmega  
     \end{array}\right),\\
     \frac{\partial \bfQ_{1n}(\boldsymbol{\beta}_0,{\bf0},\boldsymbol{\theta}^\ast)}{\partial\boldsymbol{\theta}^T}=&\frac{1}{n}\sum_{i=1}^n\frac{ \partial \bfg_i(\boldsymbol{\beta}_0,\boldsymbol{\theta}^\ast)/\partial \boldsymbol{\theta}^T}{1-\boldsymbol{0}^T\bfg_i(\boldsymbol{\beta}_0,\boldsymbol{\theta}^\ast)}+o_p(n^{-1/2})\xrightarrow{p}\left(\begin{array}{c}
      \bf0     \\
      -\bfH   
     \end{array}\right),\\
%     \frac{\partial \bfQ_{2n}(\boldsymbol{\beta}_0,{\bf0},\boldsymbol{\theta}^\ast)}{\partial\boldsymbol{\beta}^T}=&\frac{1}{n}\sum_{i=1}^n\frac{\boldsymbol{0}^T \partial^2 \bfg_i(\boldsymbol{\beta}_0,\boldsymbol{\theta}^\ast)/\partial \boldsymbol{\beta}\partial \boldsymbol{\beta}^T}{1-\boldsymbol{0}^T\bfg_i(\boldsymbol{\beta}_0,\boldsymbol{\theta}^\ast)}+o_p(n^{-1/2})\xrightarrow{p}\bf0,\\
%     \frac{\partial \bfQ_{2n}(\boldsymbol{\beta}_0,{\bf0},\boldsymbol{\theta}^\ast)}{\partial\boldsymbol{\theta}^T}=&\frac{1}{n}\sum_{i=1}^n\frac{\boldsymbol{0}^T \partial^2 \bfg_i(\boldsymbol{\beta}_0,\boldsymbol{\theta}^\ast)/\partial \boldsymbol{\beta}\partial \boldsymbol{\theta}^T}{1-\boldsymbol{0}^T\bfg_i(\boldsymbol{\beta}_0,\boldsymbol{\theta}^\ast)}+o_p(n^{-1/2})\xrightarrow{p}\bf0,\\
     \widetilde{\bfS}(\boldsymbol{\beta}_0,{\bf0},\boldsymbol{\theta}^\ast)=&\frac{1}{n}\sum_{i=1}^n\frac{ \partial \bfg_i(\boldsymbol{\beta}_0,\boldsymbol{\theta}^\ast)/\partial \boldsymbol{\beta}}{1-\boldsymbol{0}^T\bfg_i(\boldsymbol{\beta}_0,\boldsymbol{\theta}^\ast)}+o_p(n^{-1/2})\xrightarrow{p}\left(\begin{array}{cc}
      -\bfS    & \ \ \ \bfJ
     \end{array}\right),
\end{align*}
we have 
\begin{align}
  \nonumber {\bf0}&= \left(\begin{array}{c}
         \bfQ_{1n}(\boldsymbol{\beta}_0,\bf0,\boldsymbol{\theta}^\ast)+\left(\begin{array}{c}
          \bf0     \\
          -\bfQ_{3n}(\boldsymbol{\beta}_0,\bf0,\boldsymbol{\theta}^\ast)     
         \end{array} \right) \\
         \bf0\\
    \end{array}\right)+\left(\begin{array}{cc}
          \frac{\partial \bfQ_{1n}(\boldsymbol{\beta}_0,{\bf0},\boldsymbol{\theta}^\ast)}{\partial\boldsymbol{\lambda}}&\frac{\partial \bfQ_{1n}(\boldsymbol{\beta}_0,{\bf0},\boldsymbol{\theta}^\ast)}{\partial\boldsymbol{\beta}}\\
         \widetilde{\bfS}(\boldsymbol{\beta}_0,{\bf0},\boldsymbol{\theta}^\ast)&\bf0
    \end{array}\right)\\&\cdot \left(\begin{array}{c}
      \widehat{\boldsymbol{\lambda}}\\
         \widehat{\boldsymbol{\beta}}_{\pi\widehat{\theta}-2}-\boldsymbol{\beta}_0  
       %  \widehat{\boldsymbol{\theta}}_\text{MLE}-\boldsymbol{\theta}_0
    \end{array}\right)
    +o_p(n^{-1/2}):=\bfv_n+\bfG_n\left(\begin{array}{c}
      \widehat{\boldsymbol{\lambda}}\\
         \widehat{\boldsymbol{\beta}}_{\pi\widehat{\theta}-2}-\boldsymbol{\beta}_0  
       %  \widehat{\boldsymbol{\theta}}_\text{MLE}-\boldsymbol{\theta}_0
    \end{array}\right) +o_p(n^{-1/2})
\end{align}
with% $\xrightarrow{d}N(0,)$
\begin{align*}
  % \sqrt{n}  \left(\begin{array}{c}
  %    \widehat{\boldsymbol{\lambda}}\\
  %       \widehat{\boldsymbol{\beta}}_\text{EL2}-\boldsymbol{\beta}_0  
       %  \widehat{\boldsymbol{\theta}}_\text{MLE}-\boldsymbol{\theta}_0
  %  \end{array}\right)
  \sqrt{n}\bfv_n\xrightarrow{d}N\left({\bf0},\boldsymbol{\Sigma}\right), 
\end{align*}
and thus 
\begin{align*}
   \sqrt{n}  \left(\begin{array}{c}
      \widehat{\boldsymbol{\lambda}}\\
         \widehat{\boldsymbol{\beta}}_{\pi\widehat{\theta}-2}-\boldsymbol{\beta}_0  
  %  \widehat{\boldsymbol{\theta}}_\text{MLE}-\boldsymbol{\theta}_0
    \end{array}\right)\xrightarrow{d}N\left({\bf0},\bfG^{-1}\bfSigma \bfG^{-T}\right)
  \end{align*}
where 
\[
\bfSigma=\left(\begin{array}{ccc}
    \bfS & -\bfU^T&\bf0 \\
    -\bfU & \bfOmega-\bfV-\bfV^T+\bfW&\bf0\\
    \bf0&\bf0&\bf0
   \end{array}
   \right);
\]
\begin{align*}
  % \sqrt{n}  \left(\begin{array}{c}
  %    \widehat{\boldsymbol{\lambda}}\\
  %       \widehat{\boldsymbol{\beta}}_\text{EL2}-\boldsymbol{\beta}_0  
       %  \widehat{\boldsymbol{\theta}}_\text{MLE}-\boldsymbol{\theta}_0
  %  \end{array}\right)
  \bfG_n\xrightarrow{p}\bfG:=\left(\begin{array}{ccc}
    -\bfS & \bf0 &-\bfS \\
    \bf0 & -\bfOmega&\bfJ^T\\
    -\bfS&\bfJ&\bf0
   \end{array}
   \right)=\left(\begin{array}{cc}
    \bfA & \bfB^T \\
    \bfB & \bf0\\
   \end{array}
   \right).
\end{align*}
where \[\bfA=\left(\begin{array}{cc}
    -\bfS & \bf0\\
    \bf0 & -\bfOmega\\
   \end{array}
   \right), \ \bfB=\left(\begin{array}{cc}
    -\bfS & \bfJ
   \end{array}
   \right).\]
Inverting $\bfG$, 
\begin{align*}
    \bfG^{-1}&=\left(\begin{array}{cc}
    *&*\\
    (\bfB\bfA^{-1}\bfB^T)^{-1}\bfB\bfA^{-1}&-(\bfB\bfA^{-1}\bfB^T)^{-1}
    \end{array}\right)\\
&=-(\bfB\bfA^{-1}\bfB^T)^{-1}\left(\begin{array}{ccc}
    *&*&*\\
    *&*&*\\
    -\bfI&\bfJ\bfOmega^{-1}&\bfI
    \end{array}\right)\\
&= (\boldsymbol{S}+\boldsymbol{J}\boldsymbol{\Omega}^{-1}\boldsymbol{J}^{T})^{-1}\left(\begin{array}{ccc}
    *&*&*\\
    *&*&*\\
    -\bfI&\bfJ\bfOmega^{-1}&\bfI
    \end{array}\right).
\end{align*}
So the asymptotic covariance matrix of $\sqrt{n}(\widehat{\boldsymbol{\beta}}_{\pi\widehat{\theta}-2}-\boldsymbol{\beta}_0)$, i.e., the bottom-right block of $ \bfG^{-1}\bfSigma \bfG^{-T}$ is 
\begin{align*}
&\ACov\{ \sqrt{n}(\widehat{\boldsymbol{\beta}}_{\pi\widehat{\theta}-2}-\boldsymbol{\beta}_0)\}\\
=&(\boldsymbol{S}+\boldsymbol{J}\boldsymbol{\Omega}^{-1}\boldsymbol{J}^{T})^{-1}\left(-\bfI\ \bfJ\bfOmega^{-1}\ \bfI
    \right)\left(\begin{array}{ccc}
    \bfS & -\bfU^T&\bf0 \\
    -\bfU & \bfOmega-\bfV-\bfV^T+\bfW&\bf0\\
    \bf0&\bf0&\bf0
   \end{array}
   \right)\left(\begin{array}{c}
    -\bfI\\\bfOmega^{-1}\bfJ^T\\\bfI
    \end{array}\right)\\
    &\cdot(\boldsymbol{S}+\boldsymbol{J}\boldsymbol{\Omega}^{-1}\boldsymbol{J}^{T})^{-1}\\
=& (\boldsymbol{S}+\boldsymbol{J}\boldsymbol{\Omega}^{-1}\boldsymbol{J}^{T})^{-1}\left\{ \boldsymbol{S}+\boldsymbol{J}\boldsymbol{\Omega}^{-1}\boldsymbol{U}+\boldsymbol{U}^{T}\boldsymbol{\Omega}^{-1}\boldsymbol{J}^{T}+\boldsymbol{J}\boldsymbol{\Omega}^{-1}(\boldsymbol{\Omega}-\boldsymbol{V}-\boldsymbol{V}^{T}+\boldsymbol{W})\boldsymbol{\Omega}^{-1}\boldsymbol{J}^{T}\right\}\\
&\cdot(\boldsymbol{S}+\boldsymbol{J}\boldsymbol{\Omega}^{-1}\boldsymbol{J}^{T})^{-1}.
\end{align*}
  which is the same as $\ACov\{ \sqrt{n}(\widehat{\boldsymbol{\beta}}_{\pi\widehat{\theta}-1}-\boldsymbol{\beta}_0)\}$.

\nocite{*} 
\bibliography{bib_twophase}
\bibliographystyle{acm}
\end{document}